\documentclass[12pt]{article}
\pdfoutput=1
\usepackage{jcapmod}
\usepackage{booktabs}
\usepackage{mathtools}
\usepackage[english]{babel}
\usepackage{amsmath,amssymb,amsbsy,amstext, amsthm,xcolor}
\usepackage{graphicx}
\usepackage{amsfonts}
\usepackage{amssymb}
\usepackage{float}
\usepackage[utf8]{inputenc}
\RequirePackage{color}

\usepackage{colortbl}
\definecolor{green2}{cmyk}{0, 1, 0.5, 0}
\definecolor{lightgreen}{cmyk}{0.2, 0, 0.2, 0.2}
\definecolor{lightgray}{cmyk}{0.1,0.2,0,0.1}
\definecolor{lightgray2}{cmyk}{0.4,0.4,0,0.8}
\definecolor{black}{cmyk}{1.0,1.0,1.0,1.0}

\allowdisplaybreaks[1]

\usepackage{colortbl}
\definecolor{lightgreen}{cmyk}{0.2, 0, 0.2, 0.2}
\definecolor{lightgray}{cmyk}{0.1,0.2,0,0.1}
\definecolor{lightgray2}{cmyk}{0.1,0.1,0,0.1}

\makeatletter
\newlength{\apb@width}
\newcommand{\autoparbox}[2][c]{\settowidth{\apb@width}{#2}\parbox[#1]{\apb@width}{#2}}

\makeatother

\numberwithin{equation}{section}

\def\be{\begin{equation}}
\def\ee{\end{equation}}

\def\bea{\begin{eqnarray}}
\def\eea{\end{eqnarray}}

\def\Mp{M_{\rm Pl}}

\def\0{{\boldsymbol 0}}

\usepackage{setspace} 
\begin{document}

\begin{titlepage}

\setcounter{page}{1} \baselineskip=15.5pt \thispagestyle{empty}

\bigskip\

\vspace{1cm}
\begin{center}

{\fontsize{20}{28}\selectfont  \sffamily \bfseries {Fat Inflatons, Large Turns and the $\eta$-problem 
 \\}}

\end{center}

\vspace{0.2cm}

\begin{center}
{\fontsize{13}{30}\selectfont Dibya Chakraborty$^{\dagger,}$\footnote{\texttt{dibyac@fisica.ugto.mx}}, Roberta Chiovoloni$^{\ddag,}$\footnote{\texttt{roberta.chiovoloni@gmail.com}}, Oscar Loaiza-Brito$^{\dagger,}$\footnote{\texttt{oloaiza@fisica.ugto.mx}}, Gustavo Niz$^{\dagger,}$\footnote{\texttt{g.niz@ugto.mx}}, Ivonne Zavala$^{\ddag,}$\footnote{\texttt{e.i.zavalacarrasco@swansea.ac.uk}}} 
\end{center}

\begin{center}

\vskip 8pt
\textsl{$^\dagger$ Departamento de F\'isica, Universidad de Guanajuato, Loma del Bosque No.  103 Col.~Lomas del Campestre, C.P 37150 Le\'on, Guanajuato, M\'exico}\\
\textsl{$^\ddag$ Physics Department, Swansea University, SA2 8PP, UK}
\vskip 6pt

\end{center}

\vspace{1.2cm}
\hrule \vspace{0.3cm}
\noindent
 It is commonly believed  that a successful period of inflation driven by a single or several scalar fields requires a specific hierarchy of  masses given by $M_{inf}\ll H \ll M_{heavy}$, where $M_{inf}$ can correspond to several or a single {\em light} field and $M_{heavy}$ corresponds to any {\em heavy} field that might be integrated out if it satisfies suitable conditions. This is at the heart of the so called $\eta$-problem  in inflation, since large contributions to the masses of the inflatons might spoil the  slow-roll conditions required for inflation.  We show that, while this is an unavoidable conclusion  in single field inflation, in multifield inflation, {\em heavy} fields as defined above, may be {\em fully} responsible for a successful period of what we call {\em fat slow-roll inflation}. Moreover we show that in this scenario, the {\em turning rate} of the inflationary trajectory, $\Omega/H$, is larger than one. Thus, the $\eta$-problem is evaded with large turns in fat inflation. Depending on the perturbations' mass spectra, cosmological predictions will differ either slightly or largely with respect to those of the single field case. 
We illustrate this scenario in a concrete example in Type IIB string flux compactifications, where a probe D5-brane moving along the angular and radial directions in a warped throat 
drives {\em fat D5-brane natural inflation}.
An {\em instantaneous} superplanckian decay constant can be defined, consistent  with our low energy approximations.
We compute the cosmological observables, which are consistent with Planck data,   ameliorating the  tension of single field natural inflation with observations. 
 We also discuss fat inflation  in the context of recently proposed swampland de Sitter conjectures. 
 
\vskip 10pt
\hrule

\vspace{0.4cm}
 \end{titlepage}

 \tableofcontents


\section{Introduction}

Cosmological inflation \cite{Guth,Linde81,AS82}, originally proposed as a natural explanation for the homogeneity and flatness of our Universe, has been put on  firmer grounds thanks to the most recent  observations from the Planck satellite \cite{Planck2018}.
Observations are fully consistent with the
simplest inflationary scenario as the leading mechanism to account for the origin of  the anisotropies in the Cosmic Microwave Background (CMB) radiation and, thus, the formation of the large scale structures. In particular, they  agree with two robust predictions of inflation, that is, a nearly scale invariant spectrum of density perturbations and  a stochastic background of gravitational waves. 

In its simplest form, inflation is driven by a single scalar field whose potential energy dominates, driving an early  period of quasi de-Sitter accelerated expansion.  
However, recently proposed consistency conjectures \cite{Vafa1,GK,Vafa2} on the low energy effective theories derived from quantum gravity, would imply that the vanilla single scalar field inflation belongs to the set of effective theories that cannot be consistently embedded in a theory of quantum gravity, the swampland. 

On the other hand,  major  experimental efforts are  being pursued within the next decade, such as the stage four CMB-S4 experiment \cite{cmb-s4} as well as experiments aiming at detecting the B-mode polarisation in the CMB induced by primordial gravitational waves such as 
  CLASS \cite{class}, LiteBIRD \cite{Lbird}, the Simons Observatory \cite{simons}, or Probe Inflation and Cosmic Origins (PICO \cite{pico}).  Moreover, Large Scale Structure observations by forthcoming experiments, such as DESI \cite{DESI}, LSST \cite{LSST}, Euclid \cite{euclid} or SKA \cite{SKA}, may also find the existence of primordial non-Guassianities, shedding light on the driving interactions during the inflationary era \cite{Meerburg:2019qqi}.
 It is thus essential to understand, from a theoretical point of view,  what scenarios and observables  we may expect from models which go beyond the simplest single field vanilla model. In particular, multifield models of inflation are generic  in supergravity and string theory. 

We are thus in a situation where it has become essential  to move on from the vanilla single field models, both from the theoretical and experimental points of view.  In this paper we take a further step in understanding multifield inflation\footnote{Multifield inflation has been extensively studied over the last 20+ years. Thus the existing  literature is  vast and it would be impossible to include every reference in the present paper. We therefore  only refer to those papers which are most relevant for our present discussion.} in view of forthcoming experimental efforts as well as recently proposed theoretical constraints.

Our starting discussion is motivated by the following simple question: given a multiscalar Lagrangean, what are the conditions that the parameters and fields need to satisfy in order to drive a period of successful slow-roll inflation? 
We  show  that  contrary to usual belief, a long period of slow-roll inflation does not require any of the scalar fields' masses to be light (w.r.t.~the Hubble scale), that is, $M_{inf} < H< M_{heavy}$, where $M_{inf}, M_{heavy}$ correspond to the masses of one or more light inflatons and  heavy fields respectively.  On the contrary, we show that slow-roll inflation is possible when   the masses of {\em all} scalar fields are heavier than the Hubble scale. That is,  the unexpected hierarchy holds
\be\label{hierarchy}
  H \ll M_{inf}^a  \qquad  \text{ for all fields, } \,\, a=1,\dots n 
  \ee
We call this new type of inflationary attractor {\em fat inflation}  to stress the fact that it is the mass of the scalar fields themselves which is heavy (w.r.t.~the Hubble scale)\footnote{A lot of work was been done regarding the hierarchy of the fluctuations's masses, which can be classified into adiabatic and entropic. Depending on the masses of the perturbation modes, heavy fields (with respect to the Hubble scale) may, or  not, have a strong effect on the cosmological predictions \cite{Anacs1,Anacs,CAP,CP}.}. 
As we will show, {\em fat inflation} requires {\em large turning rates}, $\Omega/H\gg1$, implying a {\em non-geodesic} inflationary trajectory.
Fat inflation thus belongs to the recently discussed rapid-turn attractors \cite{Bjorkmo:2019fls}.
 Moreover,  the $\eta$-problem which arises when large contributions to the masses of the scalar fields spoil standard slow-roll inflation can be evaded in fat inflation thanks to the large turning rates\footnote{For an example of single field inflation  where the $\eta$-problem is avoided with a fat inflaton in the framework of warm inflation, see \cite{Berera:2004vm,Bastero-Gil:2019gao}.}. 

The paper is organised as follows. In section \ref{sec:1} we introduce the new fat
inflationary attractor and show that it requires large turning rates, providing a novel way to evade the $\eta$-problem. In sections \ref{sec:2} and \ref{Sec:3} we  discuss an explicit fat inflation  model in string theory, where a probe D5-brane moves along the radial and angular directions of a warped resolved conifold in a type IIB flux compactification. 
We start in section  \ref{sec:2} by introducing the  set-up, 
following the construction used in \cite{KT}. 
Next in  section \ref{Sec:3} we use the low energy action derived in  section \ref{sec:2} to construct an explicit model of fat  natural inflation. We compute the cosmological observables, which are consistent with the recent Planck data, thus improving the tension of  single field natural inflation with observations. We also include a set of parameters that gives rise to a standard hierarchy of masses and whose cosmological predictions are indistinguishable from single field. We then compute the non-linear parameter $f_{NL}$, which may help to distinguish multifield models from the single field case.  We end by discussing our findings and future directions in section \ref{disc}. We include an appendix, \ref{Models}, where we collect some field theory models in the literature with large turning rates, which happen to belong to the fat inflationary attractor.  Finally, in appendix \ref{App2} we show a set of parameters which illustrate a possible  double D-brane inflation scenario with two distinct inflationary epochs.

\section{Fat Inflatons, Large Turns and the $\eta$-problem}
\label{sec:1}

Consider a typical low energy Lagrangean for several scalar fields, which may arise from some consistent theory of quantum gravity: 
\be\label{4Daction}
S= \int{d^4x \sqrt{-\tt g} \left[\Mp^2 \frac{R_4}{2}  - \frac{g_{ab}}{2} \partial_\mu\phi^a \partial^\mu\phi^b - V(\phi^a)\right]} \,,
\ee
where ${\tt g}$ is the determinant of the four dimensional metric ${\tt g}_{\mu\nu}$, $R_4$ is the four dimensional Ricci scalar built from ${\tt g}$, while $g_{ab}$ is the metric of the scalar manifold spanned by the scalar fields $\phi^a$, with $a=1,\dots$. Although in general there can be several scalar fields, for clarity we will mostly focus on the two-field case, that is $a=1,2$. 

For cosmology we take the Friedmann-Robertson-Walker (FRW) metric 
\be\label{frw}
 ds^2 = -dt^2 + a^2(t) \,dx^idx_i \,,
 \ee
with scale factor $a(t)$, so the Hubble parameter is given by $H=\dot a/a$. 
The equations of motion thus become: 
\bea
&&H^2 = \frac{1}{3\Mp^2} \left(\frac{\dot \varphi^2}{2}  + V(\phi^a)\right) \,, \label{H} \\
&& \ddot \phi^a + 3H\dot\phi^a + \Gamma^a_{bc} \dot\phi^b\dot\phi^c + g^{ab} V_{b} =0  \,,
\label{phis}
\eea
where 
\be\label{varphi}
\dot\varphi^2 \equiv g_{ab} \dot \phi^a\dot\phi^b\,,
\ee
the Christoffel symbols in \eqref{phis} are computed using the scalar manifold metric $g_{ab}$ and $V_a$ denotes derivative w.r.t the scalar field $\phi^a$. 

We now use the common decomposition in multifield models of   tangent and normal projections of the equations above by introducing the unit tangent and normal vectors to the inflationary trajectory, $T^a$, $N^a$, as\footnote{At this point we focus on the two field case. When more fields are present, more normal vectors will be introduced.} 
\be
T^a = \frac{\dot\phi^a}{\dot\varphi}\,, \qquad T^aT_a =1\,,
\ee
and the normal is such that $N^aT_a=0$, $N^aN_a=1$.
The projected equations become
\bea
&& \ddot\varphi + 3H\dot\varphi + V_T =0 \,, \label{varphiT}\\
&& D_t T^a = - \frac{V_N}{\dot\varphi} N^a \equiv -\Omega N^a \label{Omega1}\,,
\eea
 where $V_T = V_aT^a$, 
 \be\label{Dt}
 D_tT^a = \dot T^a + \Gamma^{a}_{bc} T^b \dot \phi^c \,,
 \ee
and we introduced the dimensionful {\em turning } parameter $\Omega$, which will be important in our discussion below. 

Now, given the Lagrangean above with a given potential $V$, we would like to know what are the conditions that the potential and derivatives of the fields need to satisfy in order to drive a long period of accelerated expansion. These  are precisely  the slow-roll conditions. We now look carefully at these  and show how heavy fields can give rise to slow-roll inflation.

\subsection{Slow-Roll Fat Inflation and Large Turns}\label{SR}

Let us  analyse carefully what are the conditions that a multifield scalar theory needs to satisfy in order to drive a successful period of inflation\footnote{See  \cite{Yang} for related work.}. 
First, a nearly exponential expansion can be ensured by the requirement that the    
fractional change of the Hubble parameter per e-fold $d(\ln H)/d{\rm N}$ (where $d{\rm N}=Hdt$)  is small, that is:
\be\label{epsilonH}
\epsilon \equiv -\frac{\dot H}{H^2} = \frac{\dot\varphi^2}{2M^2_{Pl}H^2} \ll 1 \,.
\ee
Next,  inflation needs to last for a sufficiently long time so that the horizon problem is solved. This  requires that $\epsilon$ remains small for a sufficient number of Hubble times, which is measured by  the second slow-roll parameter, $\eta$:
\be\label{etaH}
\eta \equiv \frac{\dot \epsilon}{\epsilon H}  = \frac{\ddot{H}}{H \dot{H}}+2\epsilon 
= 2 \frac{\ddot{\varphi}}{H \dot{\varphi}} + 2\,\epsilon \ll 1 \,,
\ee    
Since $\epsilon \ll 1$, eq.~\eqref{etaH} implies that 
 \be\label{etaphi}
 \frac{\ddot{\varphi}}{H \dot{\varphi}}  \ll 1 \,.
 \ee
Using the Friedman equation, we can see that the  first slow-roll condition \eqref{epsilonH}, implies that $\dot\varphi^2\ll V$ and therefore, we can write
\be\label{Hslow}
H^2 \simeq \frac{V}{3M_{Pl}^2} \,.
\ee
Moreover, \eqref{etaphi} implies that we can write \eqref{varphiT} as
\be\label{phislow}
3H \dot\varphi + V_T \simeq 0 \,.
\ee
That is, the slow-roll equations to solve at the background level are \eqref{Hslow} and \eqref{phislow} and \eqref{Omega1}. 

Before proceeding, it is now useful to recall why in the single field case, the slow-roll conditions imply that the mass of the inflaton has to be much smaller than the Hubble scale, and thus the origin of the $\eta$-problem. For the single field case, we simply consider $\varphi$ as the inflaton, $V_T=V'$ and there is no third equation. The slow-roll conditions \eqref{epsilonH}, \eqref{etaH} simplify to the potential slow-roll conditions:
\be\label{slowV}
\epsilon_V \equiv \frac{M_{Pl}^2}{2} \left(\frac{V'}{V}\right)^2\ll 1\,, \qquad 
\eta_{V} \equiv M_{Pl}^2\left|\frac{V''}{V}\right| \ll1\,,
\ee
and thus the smallness of the $\eta$-parameter implies that $M_{inf}^2 \sim V'' \ll H^2$. We  now show how this conclusion is avoided in the multifield case. 

First, using  \eqref{phislow} and \eqref{Hslow}, the condition \eqref{epsilonH} implies 
\be
\epsilon_T \equiv \frac{M_{Pl}^2}{2} \left(\frac{V_T}{V}\right)^2 \ll1  \,,
\ee
that is, the tangent projection of the derivative of the potential has to be small. Next, taking the derivative of \eqref{phislow}, and imposing the condition \eqref{etaphi} making use the definitions of $D_tT^a$ and $\Omega$ in \eqref{Omega1}, \eqref{Dt}, we see that \eqref{etaphi} implies that
\be\label{VTT1}
-M_{Pl}^2\frac{V_{TT}}{V} + \frac{\Omega^2}{3H^2} + \epsilon \ll1   \,,
\ee
where $V_{TT} = T^aT^b \nabla_a\nabla_b V$ and we replaced $3H^2$ with $V$ in the first term. Since $\epsilon\ll1$, we arrive at  the first important result, that is, slow-roll (multifield) inflation implies\footnote{A similar expression appeared in footnote 9 of \cite{Christodoulidis:2018qdw} without derivation. In this paper,  large turn models were not discussed.}
 \be\label{oureq}
\left|-M_{Pl}^2\frac{V_{TT}}{V} + \frac{\Omega^2}{3H^2} \right| \ll1   \,.
\ee
Cleary  this condition can be satisfied when both terms on the left hand side are small.  However, an interesting new possibility arises when the two terms on the left hand side are large and cancel each other. This of course requires that $V_{TT}>0$. 

Let us now see what a large value of $V_{TT}/H^2$ can imply. 
Let us call  the minimal eigenvalue of the field's mass matrix, 
$\lambda$, that is 
\be\label{lambda}
 \lambda  \equiv  {\rm min}(\nabla^a\nabla_b V)\,. 
\ee
It  follows that for a unit vector  $U^a$, the following relation holds  $\lambda \leq U_a\nabla^a\nabla_b V U^b$. Taking  $U^a=T^a$, we have
\be\label{mineigen}
\lambda \leq V_{TT} \,. 
\ee
Consider now the case when  $\lambda \gg H^2$, implying that {\em all the scalar fields} are heavier than the Hubble scale. We then have 
\be\label{eigenlarge}
  H^2 \ll \lambda \\ \quad \Rightarrow   \quad H^2 \ll  V_{TT} %
  \ee
and therefore,  when $\lambda \gg H^2$, the slow-roll condition \eqref{oureq}    is satisfied when the  {\em  turning rate} is {\em large}:
\be\label{TR} 
 \frac{\Omega^2}{H^2}\gg 1\,,
\ee
which is our second important result.

Let us summarise:  the multifield slow-roll condition \eqref{oureq} can be satisfied when {\em all} the scalar fields are heavy  ($\lambda \gg H^2$) and in this case, the  turning rate $\Omega/H$ is large. 
We call this  {\em fat slow-roll inflation}, and as we show above, this type of inflationary attractor has large turning rates, $\Omega/H$. 

Notice that \eqref{oureq} implies a cancelation between $V_{TT}/V$ and $\Omega/H$, when $V_{TT}>0$. Thus, it is  possible that  $V_{TT}/V>1$  thus having  large turns, while $\lambda <0$ and small (see appendix \ref{Models}  for an example of this (AAW2)). However our point is that even when all fields are heavy, slow-roll is possible and it requires large turns\footnote{Recent multifield inflation  investigations have pointed out that small turning rates are not necessary for a successful period of slow-roll inflation \cite{Brown,Garcia-Saenz:2018ifx,Bjorkmo:2019aev,Bjorkmo:2019fls,Christodoulidis:2019jsx,Christodoulidis:2019mkj}, as we showed explicitly above. Most of these studies focus on the case of non-zero negative curvature of the scalar manifold. As we have shown above, large turning rates do not require a non-zero scalar curvature (see appendix \ref{Models} for explicit field theory examples). Moreover, as we discuss in the main text, large turning rates are possible even when the standard hierarchy of masses holds ($M_{inf}<H<M_{heavy}$), which is not fat slow-roll inflation.}.

Let us also point out that when more than two fields are present, one can define a turning rate associated to every normal direction and they will contribute to the total turning rate (see appendix \ref{Models} for an example (APR)). In appendix \ref{Models} we 
outline the construction of the simplest field theory model for two fields leading to the fat inflaton attractor with large turning  rate and we present  a collection of field theory models with large and small turning rates that have been discussed in the literature and demonstrate that   those with large turns belong to the fat slow-roll class. 
 
Note  that large  turning rates $\Omega/H$ do not imply large dimensionful turns, $\Omega$. Indeed, since $\Omega$ has dimensions of mass, it is measured in Planck units and thus we  expect $\Omega\lesssim \Mp$ in a consistent model.
Let us finally note that the geodesic displacement is measured by $|D_tT|=0$. The departure from a geodesic can be thus  measured by the dimensionless $\Omega/H$ through $|D_{\rm N} T|= \Omega/H$, where we have changed to derivatives w.r.t.~the number of efolds $d {\rm N} =Hdt $. We therefore see that fat inflation trajectories follow highly non-geodesic trajectories. Moreover, geodesic inflationary trajectories require very small  turning rates $\Omega/H\ll 1$. In table \ref{tab:5} in  appendix \ref{Models} we list a multifield inflationary example of this type (racetrack).  

 \smallskip
 
 \subsubsection*{Dynamics of the linear  perturbations.}

In multifield inflation, it is standard to decompose  the linear perturbations   in terms of  the  adiabatic  and  entropic  modes $Q_T$, $Q_N$, defined  as  the  projection  of the field fluctuations $Q^a$ in spatially flat gauge \cite{SS,GWBM,GNvT,LRP}. 
The dynamics of the primordial linear perturbations about the inflationary background for the adiabatic and entropy modes is given by the equations \cite{SS,GNvT,LRP}:
\bea
&&\ddot Q_T + 3H\dot Q_T + \left(\frac{k^2}{a^2} +m_T^2  \right) Q_T = 
\left(2\Omega Q_N\right)^{\Large\dot{}}-\left(\frac{\dot H}{H} + \frac{V_T}{\dot\varphi} \right) 2\Omega Q_N\,, \label{QT} \\
&& \ddot Q_N + 3H\dot Q_N + \left(\frac{k^2}{a^2} +M^2  \right) Q_N =- 2\Omega 
\frac{\dot \varphi}{H}\dot {\cal R}   \label{QN}
\eea
where $Q_T=T_i Q^i $, $Q_N=N_i Q^i$, $Q^i$ are the field fluctuations in spatially flat gauge, ${\cal R}$ is the comoving curvature perturbation and it is directly proportional  to the adiabatic fluctuation: 
\be
{\cal R} = \frac{H}{\dot \varphi} Q_T\,.
\ee
The adiabatic mass squared $m_T^2$ is given by 
\be\label{amass}
\frac{m^2_T}{H^2}  \equiv -\frac{3}{2}\eta- \frac{1}{4}\eta^2 -\frac{1}{2} \epsilon\eta-\frac{1}{2}\frac{\dot\eta}{H} \,,
\ee
and the entropy mass $M$ is given by 
\be\label{emass}
\frac{M^2}{H^2} = \frac{V_{NN}}{H^2} + \Mp^2\, \epsilon  \,{\mathbb R} -  \frac{\Omega^2 }{H^2}\,,
\ee
where $V_{NN} = N^iN^j \nabla_i\nabla_j V$ and ${\mathbb R} $ is the scalar manifold's Ricci scalar. At superhorizon scales, \eqref{QN} becomes
\be
 \ddot Q_N + 3H\dot Q_N + \left(M^2 +4\Omega^2 \right) Q_N \approx 0\,,
\ee
and one can define an effective entropy mass as $M_{eff}^2 = M^2 + 4\Omega^2$. The relative size of this mass scale, plays also an important role as it is related to the speed of sound for the adiabatic perturbations via the relation \cite{Anacs1,CAP,Anacs}
\be\label{cs}
c_s^{-2} =\frac{M_{eff}^2}{M^2}\,.
\ee
The dynamics of the linear perturbations and cosmological predictions will depend on 
the  hierarchies of the adiabatic and entropy modes' masses relative to each other, the Hubble parameter and  the turning rate $\Omega$. 
The curvature of the scalar manifold  ${\mathbb R} $ may also play an important role if negative and large, as it may trigger  geometric destabilisation of the entropy modes \cite{RPT}.

Notice that the adiabatic mode will  be light (w.r.t.~$H$) as long as slow-roll is satisfied (see \eqref{amass}), which is the case in the fat field inflation scenario we are discussing. On the other hand, the mass of the entropic mode  will depend on the size of $\Omega/H$, the curvature of the scalar manifold ${\mathbb R}$ and $V_{NN}/H^2$. 
For example, if besides $M \gg H$, the hierarchy $M_{eff}\gg M$ holds, the speed of sound \eqref{cs} can be reduced, with observable consequences \cite{Anacs1,Anacs,AAW}. Other possibilities can arise  as  discussed in sidetracked inflation \cite{Garcia-Saenz:2018ifx} and orbital inflation \cite{Orbital,Orbital2} where the mass of the entropic modes is (much) smaller than $H$.  

Let us  see what possibilities may arise in the heavy inflation model. Note first that we can take $N^a$ as a unit vector instead of $T^a$ as we did above to write an analogous inequality to \eqref{mineigen} in terms of $V_{NN}$, that is $\lambda\leq V_{NN}$. 
 Imposing   \eqref{eigenlarge}  also implies that $H^2\ll V_{NN}$,
which could  dominate or not over the other terms in the entropic mass \eqref{emass}. 

If the scalar manifold curvature is  negative and very large, $M$ may in principle become  small or even tachyonic. On the other hand, note that for the effective entropic mass to be much larger than $M$, ($M_{eff}\gg M$)  thus having a smaller than unity speed of sound, it is necessary that $\Omega^2$ be larger than $M^2$, which implies that $5\Omega^2 \gg V_{NN} + H^2\epsilon{\mathbb R}$.  

\subsubsection{Fat  Inflation and the $\eta$-Problem}

Let us briefly comment on the relevance of the heavy field inflationary attractor we have discussed for the so called $\eta$-problem. As we have shown, fat inflation has the unusual hierarchy of masses $H \ll M_{inf}$, where $M_{inf}$ corresponds to the mass of the ``lightest" field driving inflation. In the standard lore, such hierarchy of masses cannot drive a period of successful inflation, since large contributions to the masses of the inflatons might spoil the required flatness and therefore slow-roll conditions required for inflation. However, we have seen that fat inflation works with large masses when the turning rates are large. Therefore,   previous statements on inflation bases on light  inflatons  need to be revisited
In particular, in supergravity inflationary constraints were discussed long ago in  \cite{Marta1}, assuming the need for light fields. We leave for future work a detailed analysis of these constraints and more generally of fat inflation  
and large turns  in supergravity.

\subsection{Fat  Inflation and the Swampland}\label{sec:swamp}
We conclude this section by  making a connection between fat  inflation and the recently proposed dS conjectures\footnote{One should keep in mind that these conjectures have not been proved, and should therefore be considered with care.}  \cite{Vafa1,GK,Vafa2}, which require that 
\bea
 \frac{\nabla V}{V} &\geq& \frac{c}{\Mp}  \qquad {\text{or}} \label{swamp1}  \\
  \frac{{\rm min} (\nabla^a \nabla_b V)}{V} &\leq& -\frac{c'}{\Mp^2}  \label{swamp2}
\eea
where $\nabla V \equiv \sqrt{g^{ab}V_aV_b}$
and $c, c'$ are some ${\cal O}(1)$ constants.
It was shown in \cite{AP1} that in multifield inflation, the first condition can be satisfied, so long as the  turning rate $\Omega/H$  is sufficiently large. This can easily be seen as follows. Generalising the potential slow-roll parameter \eqref{slowV} to the multifield case we have 
\be
\epsilon_V \equiv \frac{\Mp^2}{2} \frac{V^aV_a}{V^2} = \epsilon_T + \frac{\Omega^2}{9H^2}\epsilon \,,
\ee
that is:
\be\label{epsV1}
\epsilon_V = \epsilon\left(\frac{\epsilon_T}{\epsilon}   + \frac{\Omega^2}{9H^2}\right)\,.
\ee
When $\epsilon_T\simeq \epsilon$,  one arrives at the relation presented in \cite{AP1,HP}:
\be\label{epsV2}
\epsilon_V \simeq \epsilon\left(1  + \frac{\Omega^2}{9H^2}\right)\,,
\ee
and therefore, one sees that in a multifield inflationary model, where $\Omega\ne 0$, for sufficiently large turning rate  $\Omega/H$ (and suitable values of $\epsilon$),  $\epsilon_V$ can be of order one\footnote{Since there is no calculation of the constant $c$ an order one parameter can fall in a large range of values.}. 

However, \eqref{epsV2} does not tell us how to achieve large turns given a multifield model of inflation. We have provided an answer above in eq.~\eqref{eigenlarge} and \eqref{oureq}: in order to get large turns, a sufficient condition is to  consider models where 
\be
H^2\ll \lambda  \leq V_{TT}\,, 
\ee
that is, multifield {\em fat field inflation}. Clearly in this case, the second condition \eqref{swamp2} is not satisfied.  

Let us also comment on another conjecture, the Distance Swampland Conjecture (DSC)  \cite{SDC}. Roughly, it claims  that the  geodesic displacement between two points in field space is bounded, again by an order one number in Planck units, that is:
\be\label{SDC}
\Delta \phi \lesssim \tilde c \, M_{Pl} \,,
\ee
with $\tilde c\sim {\cal O}(1)$. Otherwise a tower of light states emerges 
which would  spoil the low energy effective description. 
A recent discussion on multifield inflation and the DSC has appeared in  \cite{BPR}. So here we simply  stress that   inflationary trajectories  with large  turning rates $\Omega/H\gtrsim 1$  differ strongly from a geodesic and thus \eqref{SDC} does not apply. Moreover, an almost geodesic trajectory requires a very small turning rate value $\Omega/H\ll 1$. (See appendix \ref{Models} for a concrete example).

In the next two sections we discuss an explicit example of of fat inflation where a probe D5-brane moves along the angular and radial directions of a warped resolved conifold in a type IIB string theory compactification.

\section{D5-brane Inflation supergravity set-up}\label{sec:2}

In this section we  present the supergravity set-up where we study a concrete example of  fat D5-brane inflation. In the next section we will use the results discussed here to study the full cosmological evolution and predictions of this model.

Consider a flux compactification of type IIB string theory on an orientifold  Calabi-Yau threefold \cite{GKP}, where the use of internal fluxes generates a warped throat in the internal space. 

The low energy 10D action of type IIB  supergravity, together with local sources in the Einstein frame, is given by
\bea\label{eq:3}
S_{IIB}&=&-\frac{1}{2{\kappa}^2_{10}}\int_{M_{10}}d^{10}x\sqrt{|g|}\left(\!R-\frac{|\partial\tau|^2}{2({\rm Im}\,\tau)^2}-\frac{|G_3|^2}{2{\rm Im}\,\tau}-\frac{|\tilde{F}_5|^2}{4\cdot5!}\right) \nonumber \\
&+&\frac{1}{8i{\kappa}^2_{10}}\int_{M_{10}}\frac{C_4\wedge G_3\wedge\bar{G}_3}{{\rm Im}\,\tau}+S_{loc}\,,
\eea
where $\tau=C_0+ie^{-\phi}$ is the axio-dilaton and the three-form flux, $G_3=F_3-\tau H_3$, is a combination of the Ramond-Ramond  (RR) and Neveu-Schwarz--Neveu-Schwarz (NS-NS) three-form fluxes: $F_3 = dC_2$, $H_3 = dB_2$ and $  \tilde{F}_5=F_5-\frac{1}{2}C_2\wedge H_3+\frac{1}{2}B_2\wedge F_3$. Note that $F_5$ is self-dual five-form with $\tilde{F}_5=\star_5\tilde{F}_5$. The 10D gravitational constant is given by $\kappa^2_{10}=\frac{1}{2}(2\pi)^7 g_s^2 \alpha'^4$, where $\sqrt{\alpha'}=\ell_s$ is the string length  and $g_s=e^{\langle \phi\rangle}$. To add $Dp$-branes into the setup, we include in the local action $S_{loc}$, a DBI term plus a Chern-Simons contribution, namely
\be\label{Sloc}
S_{loc}=S_{DBI}+S_{CS}\,.
\ee

We consider a warped metric ansatz for a flux compactification given by \cite{GKP},
\begin{equation}\label{eq:5}
ds^2=e^{2A(y)}{\tt g}_{\mu\nu}dx^{\mu}dx^{\nu}+e^{-2A(y)}\tilde{g}_{mn} dy^m dy^n \,,
\end{equation}
where the warp factor $A(y)$ and the unwarped internal metric, denoted by $\tilde{g}_{mn}$, depend only on the internal six-dimensional coordinates $y^m$ and the maximally symmetric 4D spacetime has metric ${\tt g}_{\mu\nu}$.
The self-dual $\tilde{F}_5$  takes the form \cite{GKP},
\begin{equation}\label{eq:6}
\tilde{F}_5=(1+\star_{10})d\alpha(y)\wedge\sqrt{-{\rm det} {\tt g}_{\mu\nu}} \,dx^0\wedge dx^1\wedge dx^2\wedge dx^3 \,,
\end{equation}
where $\alpha(y)$ is a function of the internal coordinates. 
The 10D Einstein equations and the 5-form Bianchi identity imply \cite{ref35} 
\begin{equation}\label{eq:8}
\tilde{\nabla}^2\Phi_{-}=R_4+\frac{e^{8A(y)+\phi}}{24}|G_{-}|^2+e^{-4A(y)}|\partial\Phi_{-}|^2+ {\rm local}
\end{equation}
where $ R_4 $ is the four-dimensional Ricci scalar. This curvature term is not present when the 4D spacetime is taken to be Minkowski \cite{GKP}, but in the case of inflation, this spacetime is quasi-de Sitter, hence 
$R_4 \simeq 12H^2 $, with $H$  the Hubble parameter. Furthermore, the  Laplacian $\tilde \nabla$ is constructed from the unwarped internal metric $\tilde g_{mn}$, and we define the following fields
\begin{equation}\label{PhiG}
\Phi_{-}\equiv e^{4A(y)}-\alpha(y), \qquad  G_{-}\equiv*_6G_3-iG_3 \,.
\end{equation}

Integrating \eqref{eq:8} over the internal space in the case $R_4=0$ (assuming no boundary contribution at infinity) the LHS vanishes as it is a total derivative. Since each term on the RHS is positive semi-definite, each must individually vanish at leading order, giving the imaginary self-dual (ISD) solution $G_-= 0$ and $\Phi_-= 0$. 

In order to construct cosmological solutions, we start in the non-compact limit with an infinitely long warped throat, supported by the ISD flux solutions $G_-= \Phi_-=0$. To obtain dynamical 4D gravity, we then cut off the warped throat at some large radial distance, $r_{UV}$, and glue it to a compact bulk Calabi-Yau (CY). While the full metric on the bulk is not known, the metric on the warped throat is explicitly known for certain cases, such as the one used here, corresponding to the well known resolved conifold (RC) \cite{Candel,RCmetric,Kleb}. Given that we partially know the full metric, we only consider the possibility of having inflation well within the warped throat region. 
Perturbations to $\Phi_-$ arise as a result of this gluing procedure and are solutions to the Poisson equation \eqref{PhiG}.  Assuming that the gluing procedure induces  small corrections to $\Phi_-$,  and  $G_-$ of the same order, the leading order perturbation of $\Phi_-$ in the large throat limit is a solution to the homogeneous Laplace equation: 
  \be\label{Poisson1}
  \tilde \nabla^2 \Phi_h =0 \,,
  \ee
  while $\overline{\Phi}_-$ is  the solution to the Poisson equation arising when we consider the effect of a non-negligible $R_4$: 
  \be\label{Poisson2}
\tilde  \nabla^2 \overline{\Phi}_-= R_4\,, 
  \ee
The solutions to \eqref{Poisson1} and \eqref{Poisson2} depend on the unwarped internal 6D geometry and were computed in \cite{KT} for the  RC geometry. These will be relevant for the potential for the D5-brane positions and will be  presented  in subsection \ref{ssec:2}.

\subsection{The warped resolved conifold}

We now consider the warped resolved conifold (WRC)  \cite{RCmetric,Kleb} where we study the dynamics of moving probe D-branes. This resolved conifold (RC) is one of the two smooth versions of the non-compact Calabi-Yau threefold, the conifold  \cite{Candel}, which is  a cone over the base  $T^{1,1}=\frac{SU(2)\times SU(2)}{U(1)}$, which can be thought topologically as  an $S^2\times S^3$.   
At the tip of the cone, the volume of both spheres vanishes, and there is a singularity. This can be removed by either deformation or resolution.  In the case of deformation the $S^2$ sphere of the $T^{1,1}$ base shrinks at the tip and it takes the shape of a $S^3$  giving rise to deformed conifold (DC). In the case of resolution the singularity is removed by blowing up the two-sphere of  the $T^{1,1}$ giving rise to the resolved conifold \cite{Candel,Kleb}.
The warped 10D spacetime is  obtained by placing a stack of $N$ D3-branes at the tip of the RC, extended along the four non-compact spacetime directions localised at the north pole of the $S^2$ at the tip of the RC. 
Since localising the stack at the north pole specifies an angle, the warp factor  has both angular and radial dependence\footnote{In contrast,  the warp factors  depend only on the radial coordinate in the case where the internal geometry is the singular or deformed conifold, and is an assumption usually made for generic warped throats.}.
The resulting geometry is the  resolved conifold with 10D metric \cite{RCmetric,Kleb}
\begin{equation}\label{wrc1}
ds^2= {\mathcal H}^{-1/2}(\rho,\theta_2)ds^2_{FRW}+  {\mathcal H}^{1/2}(\rho,\theta_2)ds^2_{RC}\,,
\end{equation}
where we  take the 4D spacetime to be FRW for our cosmological application, and the 6D unwarped space is the RC,  whose metric is given by \cite{RCmetric}
\bea\label{rc}
ds^2_{RC}=\tilde{g}_{mn}dy^m dy^n&=&\bigg(\frac{r^2+6u^2}{r^2+9u^2}\bigg)dr^2+\frac{1}{9}\bigg(\frac{r^2+9u^2}{r^2+6u^2}\bigg)r^2(d\psi+\cos\theta_1 d\phi_1+\cos\theta_2 d\phi_2)^2 \nonumber \\
&+&\frac{1}{6}r^2(d\theta_1^2+\sin^2\theta_1 d\phi_1^2)+\frac{1}{6}(r^2+6u^2)(d\theta_2^2+\sin^2\theta_2 d\phi_2^2)\,,
\eea
here  $u$ is the resolution parameter. It is also the natural length scale of the resolved conifold. We have also defined the dimensionless coordinate $\rho=r/3u$. 
The warp factor, ${\mathcal H}(\rho,\theta_2)$ is the solution to the Green's function equation for the Laplace operator on the RC. An exact expression for the WRC warp factor is given by \cite{Kleb}
\be\label{WF1}
\mathcal{H}(\rho,\theta)=(L_{T_{1,1}}/3u)^4\sum_{l=0}^{\infty}(2l+1)H^A_l(\rho)P_l[\cos\theta]\,,
\ee
the length scale of the $T^{1,1}$ is set by $L_{T^{1,1}}^4 = (27/4) \pi N g_s \ell_s^4$; $P_l$ are the  Legendre polynomials, and the radial functions $H_l^A(\rho)$ are given in terms of the $_2F_1(a, b, c; z)$ hypergeometric functions as
\bea\label{HA}
H_l^A(\rho) = \frac{2\Gamma(1+\beta)^2}{\Gamma(1+2\beta)}{\rho^{2+2\beta}} _2F_1(\beta, 1+\beta, 1+2\beta; -1/\rho)\,,
\eea
where $\beta = \sqrt{1+(3/2)l(l+1)}$. For the cosmological solutions we study below, we  keep only the $l=0$ mode in \eqref{WF1}, which corresponds to the 'smeared' solution in \cite{RCmetric}. Taking the limit of small $r$, one can see that the apparent singularity of this mode is removed once the full sum is considered \cite{Kleb}.

\subsection{Moving D-branes in  the warped throat}\label{ssec:2}

We now consider a D5-brane extending in the four  non-compact dimensions, wrapping $p$-times a 2-cycle in the internal WRC space and moving along the radial and one angular direction in the compact space. We follow closely   \cite{KT}, 
where the authors focused on  the potential for the angular direction,  obtaining a superplanckian decay constant to realise a model of {\em single field natural inflation}. 
As in \cite{KT}, we turn on a non-zero electric flux on the worldvolume of the D5-brane, $F_2$, which generates a non-trivial cosine contribution to the potential for the angular direction, as we review below. 

In the next section we will study in detail the cosmological evolution  for the two field inflationary evolution using the full potential computed in \cite{KT} for the radial and angular coordinates. The potential can in principle  support either single field inflation but more interestingly a multifield evolution, with both fields moving during  inflation giving rise to either fat  or standard inflation with large and small turning rates respectively. A  double D-brane inflation can in principle also be realised. The radial field drives a first period of inflation, relevant for the CMB scales; inflation then stops briefly until the  angular field  drives a second period of accelerated expansion, which might be interesting for phenomenological applications such as the production of primordial black holes\footnote{We present an example of this in appendix \ref{App2}. Although the model is unrealistic from the cosmological and theoretical points of view. }. In section \ref{Sec:3} we focus on a set of parameters which gives a fat natural inflation model with  large turning rates. We also present a set of parameters which give rise to a standard type of inflation with small turning rates.

Before looking into the cosmology, in this subsection we  review the  action describing the D-brane dynamics and the derivation of the scalar potential computed in \cite{KT}. 
The D5-brane dynamics are described by the DBI and CS terms:
\bea\label{5action}
S_5 = && S_{DBI_5} + S_{CS_5} \nonumber \\
= && -T_5 \int_{{\cal W}_6}{d^6\xi  \sqrt{-{\rm det} (P_6\left[g_{ab} +B_{ab}+ 2\pi\alpha'F_{ab}\right])}}  
\nonumber \\
&&  \hskip3cm + \mu_5  \int_{{\cal W}_6}{P_6\left[ C_6 + C_4\wedge(B_2+2\pi\alpha' F_2)\right]}\,, 
\eea
where 
\be
\mu_5 = \left[ (2\pi)^5 \ell_s^6 \right]^{-1} \,, \qquad {\text {and} }  \qquad T_5 = \mu_5 g_s^{-1}, 
\ee
$F_2$ is the world volume gauge field, $B_2$ is the NSNS 2-form field pulled back on the brane   and $P_6$ is the pullback of  a 10D tensor to the six dimensional  brane worldvolume 
 
We take the simple embedding of the D5-brane in the 10D spacetime as in \cite{BLS,KT}: 
\be
\xi^a = (x^\mu,\theta_1,\phi_1) \,,
\ee
where $\mu=0,1,2,3$  are the non-compact coordinates.  The wrapping of the brane of the 2-cycle $\Sigma_2$ in   the  internal space  is specified by the natural 2-cycle in $T^{1,1}$, given by 
\be
r={\rm const.} \,, \qquad \psi ={\rm const.} \,, \qquad \theta_2 = f(\theta_1) =-\theta_1\,, \qquad \phi_2 = g(\phi_1) = -\phi_1\,.
\ee
 Having specified the embedding and wrapping, we can now compute the pullback of the 10D metric $g_{MN}$ defined as
 \be
 P_6[g]_{ab}=\frac{\partial x^M}{\partial\xi^a}\frac{\partial x^N}{\partial\xi^b} g_{MN}\,,
 \ee
which gives us the induced  metric on the brane, with components:
\bea
P_6[g]_{00} &=&  -{\cal H}^{-1/2}(1-{\cal H} v^2)\,,  \\
P_6[g]_{ij}&=&  a^2 {\cal H}^{-1/2}  \delta{ij} \,,  \\
P_6[g]_{\theta_1\theta_1}&=&  \frac{1}{3}{\cal H}^{1/2} (r^2+3u^2) \,,\\
P_6[g]_{\phi_1\phi_1}&=&   \frac{1}{3}\sin^2\theta_1{\cal H}^{1/2} (r^2+3u^2) \,.  
\eea
We will be considering  the D5-brane to be moving  along the radial and one angular direction, $\theta_2$, while it is assumed to be fixed along the other two internal dimensions.  In this case, the  speed squared of the brane is given by  
\bea\label{Vel}
v^2 &=& g_{mn} \,\dot y^m \dot y^n  = g_{rr} \dot r^2 + g_{\theta_2\theta_2} \dot\theta_2^2 \nonumber \\
&=& \left( \frac{r^2+6 u^2}{r^2+9u^2}\right)\dot r^2 + \frac{1}{6}(r^2+6u^2)\dot \theta_2^2 \,.
\eea
As we mentioned above, we turn on a non-zero worldvolume flux $F_2$ of strength $q$, along the wrapped 2-cycle (all other components of $F_{ab}$ are set to zero), so  that its pullback has the following non-zero components
\be
P_6[F_2]_{\theta_1\phi_1} =  -P_6[F_2]_{\phi_1\theta_1}  = \frac{q}{2} \sin\theta_1 \,.
\ee
With this we have all information we need to write down the total action for the D5-brane \eqref{5action}. Notting also  that  $P_6[B_2]=0$ and  $C_6=0$, the action  becomes (expanding the square root)
\bea\label{5action2}
S_5 &=& -4\pi p \,T_5 \int{d^4x \sqrt{-{\tt g}_4} {\cal H}^{-1} {\cal F}^{1/2} \left[1- \frac{1}{2}{\cal H} v^2\right] }  + 4\pi^2 \alpha' pq \,\mu_5 \int{d^4x\sqrt{-{\tt g}_4} \,\alpha (y)} \nonumber \\
&=&  \int{d^4x \sqrt{-{\tt g}_4} \left[ \frac{1}{2}g_{ij} v^i v^j   - V(r,\theta_2)    
\right]} \,,
\eea
where we used \eqref{eq:6} and \eqref{PhiG}, and we defined: 
\bea\label{functions1}
&& g_{ij} = 4\pi p T_5 {\cal F}^{1/2} {\rm diag} \left(\frac{r^2+6u^2}{r^2+9u^2}, \frac{1}{6} (r^2+6u^2) \right) \,, \qquad 
                v^i =(\dot r, \dot \theta_2)\,, \\
 && {\cal F} \equiv \frac{\cal H}{9} (r^2+3 u^2)^2 + (\pi \ell_s^2 q)^2 \,, \\
&& V(r,\theta_2) = \varphi(r)   + \gamma \left(\overline{\Phi}_- +\Phi_h \right)\,,\qquad \qquad\qquad  \gamma = 4\pi^2\ell_s^2 p q T_5 g_s \,,\\
&& \varphi(y) = 4\pi p T_5 {\cal H}^{-1} \left[   {\cal F}^{1/2} -  \ell_s^2  \pi  q g_s  \right]\,, \\
&& {\cal H} = \left(\frac{L_{T^{1,1}}}{3u}\right)^4 \left( \frac{2}{\rho^2} - 2 \ln{\left(\frac{1}{\rho^2} +1\right)}\right)\,,
\qquad \quad  L_{T^{1,1}}^4 = \frac{27 \pi}{4} N g_s \ell_s^4\,.
\eea
Here $\Phi_-=\overline{\Phi}_-+\Phi_h $, is the  solution to the Poisson equation,  while $\Phi_h$ is the solution to the homogeneous equation \eqref{Poisson1} while  $\overline{\Phi}_-$ is the solution due to the correction of the Ricci scalar \eqref{Poisson2}. We focus on solutions of the Laplace equation which are invariant under the $SU(2)_1 \times U(1)_\psi$  which rotates the $(\theta_1, \phi_1)$ and $\psi$ coordinates of the shrinking $S^3$. 
The solutions were presented in \cite{KT} and are given by (remember that $\rho=r/3u$)
\bea
\Phi_h (\rho,\theta_2,\phi_2)= \sum_{l=0}^{\infty}\sum_{m=-1}^{m=l}\left[a_lH_l^A(\rho) + b_lH_l^B(\rho) \right] Y_{lm}(\theta_2,\phi_2) \,, \label{Phih1} \\
\overline{\Phi}_- =  \frac{5}{72}\left[ 81\left(9\rho^2 -2 \right)\rho^2 + 162 \log{(9\left(\rho^2+1\right))} -9 -160\log(10)\right]\,,\label{Phibar}
\eea
where $(l,m)$ denote the other $ SU(2)_2$  quantum numbers of  the corresponding isometries of $T^{1,1}$. The independent solutions are given by  $H_l^A(\rho)$ in \eqref{HA}  and 
\be
H^B_l(\rho) = _2F_1(1-\beta,1+\beta,2,-\rho^2)\,.
\ee 
We refer to \cite{KT} for further details.  The  homogeneous solution $\Phi_h$ is independent of the choice of probe brane and it is valid everywhere within the WRC throat, in particular near the tip. The coefficients $a_l, b_l$ are undetermined, but small. %
 We  keep two  independent solutions (depending only on $\theta_2$)  to the Laplace equation for $(l,m)=(0,0), (1,0)$, so that $\Phi_h$ is given by\footnote{We take $b_0=0$, as this term multiplies $H_0^B=1$ and thus gives a small constant contribution to the potential given by $\lambda b_0$.}
  \bea\label{Phih}
 \Phi_h &=& a_0 \left[\frac{2}{\rho^2} - 2 \log \left(\frac{1}{\rho^2}+1\right)\right]  + 
 2 a_1  \left[6+\frac{1}{\rho^2} -2(2+3\rho^2) \log\left(1+\frac{1}{\rho^2}\right)\right] \cos \theta_2  \nonumber \\
 && \hskip7cm +  \frac{b_1}{2}\left(2+3\rho^2\right)\cos\theta_2 \,,
 \eea 
 where again, the coefficients $a_0, a_1,b_1$ are small. In \cite{KT} $a_1$  was taken to be zero. However we will keep it in our analysis of the inflationary solutions in the next section. %
 
\begin{figure}[H]
  \centering
    \includegraphics[width=0.45\linewidth]{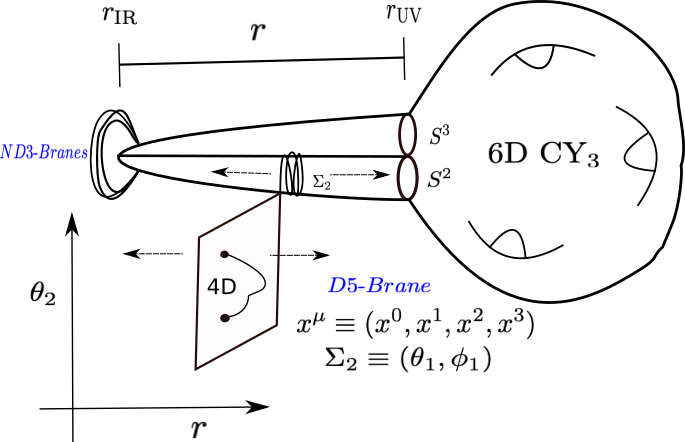}
\caption{A cartoon representation of the D5-brane embedding  in the WRC.}
\end{figure}\label{Fig1}

\subsection{Moduli stabilisation}

 We are using  the open string moduli associated to the position of a moving probe D5-brane to drive inflation and are thus intrinsically assuming that all closed string moduli, complex structure, dilaton and K\"ahler moduli, have been stabilised and are fixed at their minima. We briefly outline  how this  assumption can be realised, as discussed also in \cite{KT}, but we do not attempt to implement a full closed string stabilisation mechanism in detail in the present paper. %
 
In type IIB flux compactifications, closed string moduli are partially stabilized by turning on suitable RR and NSNS fluxes \cite{GKP}. This can be seen  in a supergravity ${\cal N}=1$ description by the scalar potential induced from the  Gukov-Vafa-Witten (GVW) superpotential in type IIB string theory ${\cal W}=\int G_3\wedge \Omega$, where $\Omega$ is the holomorphic $(3,0)$-form of the internal manifold and $G_3$ is the three-form flux defined above.  The GVW  scalar potential depends on the complex structure and the axio-dilaton moduli which can thus be stabilised by the presence of the 3-form flux $G_3$ while the K\"ahler moduli remains unfixed. Stabilisation of the K\"ahler moduli can be achieved by considering   non-perturbative corrections to the superpotential such as gaugino condensation from wrapped D7-branes. 

We now  assume that at the inflationary scale, the stabilisation of all the  closed string moduli has been completed and these moduli do not affect the brane dynamics on the WRC.  The RC has Hodge numbers $h^{2,1}=0$, $h^{1,1,}=1$ \cite{Aganagic:1999fe} meaning that there is no complex structure moduli, while there is a single K\"ahler modulus.  %
Since $h^{2,1}=0$, it is not possible to turn on $(2,1)$-form fluxes, which can preserve ${\cal N}=1$ supersymmetry. 
The axio-dilaton can be stabilsed by turning on non-supersymmetry preserving $(3,0)$-form fluxes, allowed by the existence of the non-trivial 3-cycles (the third betti number is $b_3=2(1+h^{2,1})=2$). 
When the WRC is glued to a compact CY, the number $h^{(2,1)}$ may be modified, allowing the stabilisation of the complex structure (which determines some geometry of the compact space far way from the throat) and the axio-dilaton using supersymmetry preserving $(2,1)$-fluxes.  
In summary, we expect that flux stabilisation of closed string moduli can be achieved when the WRC is glued to a compact CY without affecting the subsequent open string inflationary evolution.

\subsection{Backreaction constraints}\label{sec:back}

We now discuss briefly the constraints   on the wrapping number $p$ and  brane flux $q$ due to the D5-brane  backreaction  onto the background geometry (see also \cite{BLS,KT,KPZ}).

A D5-brane could in principle alter the warp factor and internal geometry and introduce a non-trivial profile for the dilaton. However, if its contribution to the Einstein equations is much smaller than that of the stack of the $N$ D3-branes sourcing the warped throat, then one can safely consider the D5-brane as a probe. We can estimate the size of the D5-brane contribution compared to the $N$ D3-branes as follows.
Consider the local contribution from a Dp-brane to the traced Einstein's equation, which goes as \cite{BLS,GKP}
\be
(T^m_m -T^\mu_\mu)^{\rm loc} = (7-p) T_p \,\Delta^{(9-p)}(\Sigma_{p-3}) \,,
\ee
where $\Delta^{(9-p)} (\Sigma_{p-3})=\delta^{(9-p)}(\Sigma_{p-3})/\sqrt{{\rm det} g_{9-p}} $ is the covariant delta function on the wrapped (p-3)-cycle, $\Sigma_{p-3}$. The condition that the backreaction of the wrapped D5-brane be negligible can be then written as 
\be\label{Backp1}
\frac{p}{2N}\frac{T_5}{T_3} \frac{\Delta^{(4)}(\Sigma_{2})}{\Delta^{(6)}(\Sigma_{0})} \ll 1 \,,
\ee
Using the WRC metric we find  \cite{KPZ}
\be\label{Backp2}
p\ll  \frac{T_3}{T_5} \frac{ 2N{\cal H}^{-1/2} }{ r^2 \sin\theta_1}  = \frac{12N (2\pi)^2 {\cal H}^{-1/2}}{\sin\theta_1} \frac{\ell_s^2}{r^2}\,.
\ee
This depends on the inverse of the warp factor and thus the r.~h.~s.~of \eqref{Backp2} will be smaller at the minimum $r_{min}$. As we will see, it is easy to achieve a successful period of inflation for a wide range of values of $p$ consistent with \eqref{Backp2}.
We can similarly find a bound for the brane flux $q$ by noting that it  induces a D3-brane charge due to  the CS term of the D5-brane action \eqref{5action}. Therefore it contributes to  the five form Bianchi identify as \cite{KT}:
$T_5 \rho_3^{pq\,D5} $ which should be small compared to the D3-brane contribution $T_3\rho_3^{N\,D3}$, so we require, similar to \eqref{Backp1} that  
\be\label{Backq1}
\frac{T_5 \rho_3^{pq\,D5}}{T_3\rho_3^{N\,D3}}  \ll 1\,.
\ee
Here \cite{KT}
\be
 \rho_3^{N\,D3} =  N \frac{\delta^{(6)}(\Sigma_0)}{\sqrt{{\rm det} \,g_6}} \,, \qquad
\rho_3^{pq\,D5} = p\, q  (\pi\alpha') \sin{\theta_1}\frac{\delta^{(4)}(\Sigma_2)}{\sqrt{{\rm det} \,g_6}}\,.
\ee
Therefore we arrive at the constraint 
\be\label{Backq2}
pq \ll \frac{T_3}{T_5} \frac{N}{\pi \ell_s^2\sin\theta_1} = \frac{4\pi N}{\sin\theta_1}\,.
\ee
Therefore, once we choose a value for $p$ that satisfies \eqref{Backp2}, we need to choose $q$ such that \eqref{Backq2} holds. As we will see there is a large parameter space where these conditions can be satisfied, giving rise to a successful period of inflation with large and small turning rates. 


\section{Fat D5-brane inflation in the  warped resolved conifold}\label{Sec:3}

We  now have all we need to study explicitly  the multifield D5-brane inflationary evolution, where a probe D5-brane moves inside the WRC along the radial and an angular directions: $(r,\theta)$ (from now on, we drop the subindex 2 in the angular coordinate).  Due to the complexity of the system, we solve all equations numerically. 

\subsection{Effective 4D action and cosmological equations}

Our starting action is given by (see eq.~\eqref{5action2})
\be\label{4daction}
S_4=\int d^4 x \sqrt{- {\tt g}} \left[\frac{\Mp^2}{2} R_4  + \frac{1}{2}g_{ij} v^i v^j   - V(r,\theta)    \right]
\ee
where the four dimensional metric is the FRW metric \eqref{frw},
$g_{ij}$ is defined  in \eqref{functions1} and the full expression for the scalar potential is given by (see \eqref{functions1}, \eqref{Phibar}, \eqref{Phih}):
\bea\label{Pot1}
V(r,\theta) &=&  V_0 + 4\pi p T_5 {\cal H}^{-1} \left[   {\cal F}^{1/2} -  \ell_s^2  \pi  q g_s  \right] + 
\gamma\left[\overline{\Phi}_- + \Phi_h\right] \,,
\eea
where $\gamma= 4\pi^2\ell_s^2 p q T_5 g_s $ and (see \eqref{5action2},\eqref{functions1})
\bea
{\cal F} &=& \frac{\cal H}{9} (r^2+3 u^2)^2 + (\pi \ell_s^2 q)^2 \\
\overline{\Phi}_- &=& \frac{5}{72}\left[ 81\left(9\rho^2 -2 \right)\rho^2 + 162 \log{(9\left(\rho^2+1\right))} -9 -160\log(10)\right]\, \\
\Phi_h &=& a_0 \left[\frac{2}{\rho^2} - 2 \log \left(\frac{1}{\rho^2}+1\right)\right]  + 
 2 a_1  \left[6+\frac{1}{\rho^2} -2(2+3\rho^2) \log\left(1+\frac{1}{\rho^2}\right)\right] \cos \theta  \nonumber \\
 && \hskip7cm 
 +  \frac{b_1}{2}\left(2+3\rho^2\right)\cos\theta\,.
 \label{pot}
\eea
As we explained in the previous section, the coefficients $a_0,a_1,b_1$ are arbitrary, but small (in \cite{KT} $a_1=0$). 
 We have also introduced a constant piece $V_0$, which we  tune in order to downlift  the de Sitter minimum of the potential to Minkowski. The reasons behind are twofold. 
 This term  encodes any unknown physics that may shift these minima to Minkowski. For example, due to the explicit stabilisation mechanism of the closed string moduli, which we haven't included. Moreover, the recently proposed dS swampland conjectures \cite{Vafa1,GK,Vafa2} exclude dS minima in string theory, if correct, while Minkowski minima are allowed. 

 Finally, the four dimensional Planck mass, $\Mp$ after compactification  is given by (see \eqref{eq:3})
 \be\label{eq:Mp}
 \Mp^2 \gtrsim \kappa^{-2}_{10}\,{\rm Vol}\,(T^{1,1}) \int_{0}^{u}{y^5{\cal H}(y)} \sim \frac{Nu^2}{4(2\pi)^3 g_s \ell_s^4} \,.
 \ee
 where we used that ${\rm Vol}\,(T^{1,1})  = 16\pi^3/27$ and  assumed that most of the volume comes from the throat, approximating ${\cal H}\sim L^4/\rho^4$. For concreteness, for the cosmological solutions we fix $\Mp$ to the lower bound. 

 \subsubsection*{Analysis of parameters }
 
 Before looking  into the full numerical analysis of multifield  inflationary solutions to \eqref{4daction}, let us pause here to discuss the parameters' values that we consider, taking into account  our approximations.  
 First of all, for the string weak coupling approximation to be valid we need  $g_s\ll1$. Next, we require a large number of D3-branes   $N\gg1$ so that backreaction of the probe  D5-brane is under control. As we mentioned before,  in the WRC, the $u$ parameter is the natural length of the throat, so that we can take \cite{KT} $r_{UV} =u$ and it should be larger than $\ell_s$, that is $u>\ell_s$. We also need to keep in mind the hierarchy of scales that needs to be satisfied in order for our approximations to be valid during 4D inflation \cite{Bau1,PZ}. 
 That is,  $ \Mp \gtrsim M_s \gtrsim  M_{c}\gg H$, where $M_c$ is the compactification scale and $H$ is the Hubble parameter defined as $H\equiv \dot a/a$. Taking these considerations into account, we  fix the parameters $g_s, N, u$ to ensure that this hierarchy holds and  vary the  parameters $p,q$, keeping track of the backreaction constraints \eqref{Backp2}, \eqref{Backq2}. We then choose  the coefficients $a_0, a_1, b_1 (\ll 1)$ in the potential \eqref{Pot1} such that the amplitude of the scalar perturbations matches with observations. As we will see, there is a large range of values for the parameters $p,q, a_0, a_1, b_1$ giving different types of inflationary solutions, in particular, fat slow-roll natural inflation. 

As pointed out in \cite{KT} we can expect the potential \eqref{Pot1} to drive single field natural inflation once the radial coordinate is fixed to its minimum,  $r=r_{min}$ and  so long as the decay constant, $f$,  takes superplanckian values consistent with the approximations above. It  was then shown  in \cite{KPZ} that warping and wrapping  can help in obtaining superplanckian decay constants in single field inflation, consistent with the supergravity low energy approximations and the  weak gravity conjecture (WGC) \cite{WGC}, which when applied to the axion would require that the axion decay constant be subplanckian\footnote{We refer the reader to \cite{KPZ} for details on how this conclusion could be avoided in warped single field axion models.}. 

However, one may wonder  whether the fixing of $r$ to its minimum is a good approximation, and whether this field may contribute to the inflationary evolution and  give interesting observable features. 
In  subsections \ref{LT} and \ref{ST} we present explicit numerical solutions for the two-field cosmological evolution. 
Since the radial and angular fields are evolving, we define an  {\em instantaneous decay ``constant"}, $f$ for the angular variable. This parameter depends on the radial field (see \eqref{dc} below) and thus on the inflationary trajectory, which we  deviates from a geodesic (see discussion in section \ref{SR}). For suitable choices of parameters, this instantaneous decay constant  can take  superplanckian values, consistent with the supergravity low energy approximations.
Note however that in this case, a straightforward application of the WGC is not clear. We leave for future work a study  of the implications of the  WGC for time dependent  decay constants and non-geodesic trajectories. 

In particular, in subsection \ref{LT}  we present an example of a {\em fat natural inflationary} scenario  with  large turning rate $\Omega/H>1$. We show that the cosmological predictions differ from single field natural inflation and  enter the $95\%$CL regions of the latest Planck results \cite{Planck2018}. We compute the local non-Gaussianity parameter $f_{NL}$, which can help distinguish multifield models from single field. 
In subsection \ref{ST} we use  a different  choice of parameters to provide an example of a standard inflation with the usual hierarchy of masses and small  turning rate ($\Omega/H\sim 0.35$).
 Further in the appendix \ref{App2}, we show a toy model of {\em double D-brane inflation}  where the CMB scale can be fixed at the first inflationary period, while other interesting features can arise from the second period. This model however would require the brane to start moving from inside the bulk region, which lies  outside the consistency range of our approximations and therefore we do not consider it further.

\subsection*{Cosmological evolution}

We are now ready to study the D5-brane multifield inflationary evolution in the warped throat. 
The equations of motion for \eqref{4daction} in the FRW background are given by \eqref{H}, \eqref{phis}, which we rewrite here for clarity:  
\bea
&& H^2=\frac{1}{3\Mp^2}\left(\frac{\dot\varphi^{\,2}}{2} +V(\phi^j) \right)\,, \label{eqH} \\
&&\ddot\phi^{\,i}+ 3H\dot\phi^{\,i}+ \Gamma_{jk}^{i} \dot\phi^{\,j}\dot\phi^{\,k} +  g^{ij}\partial_j V =0  \,, \label{eqphis}
\eea
where  $\phi^i = (r,\theta)$, 
\be\label{varphi1}
\dot\varphi^2 = g_{ij} \dot\phi^{\,i} \dot\phi^{\,j} = 
g_{rr} \dot r^{\,2} + g_{\theta\theta}  \dot\theta^{\,2} \,,
\ee
and  the Christoffel symbols are computed with respect to the scalar metric $g_{ij}$, which we recall here 
\be\label{metricij}
g_{rr} = 4\pi p T_5 {\cal F}^{1/2} \frac{r^2+6u^2}{r^2+9u^2} \,, \qquad 
g_{\theta\theta} = 4\pi p T_5 {\cal F}^{1/2} \frac{r^2+6u^2}{6} \,.
\ee 
 
We now look at different explicit inflationary solutions. As we mentioned before, we start by presenting an explicit example of {\em fat natural inflation} with large turning rate $\Omega/H$.

\subsection{Fat D5-brane inflation with large turning rate}\label{LT}

We now present an explicit set of parameters which realises  {\em fat slow-roll inflation}  where the dimensionless turning rate $\Omega/H$ is large while  the dimensionful $\Omega$ remains small (in Planck units). 
%

We solve the full equations of motion  \eqref{eqH}, \eqref{eqphis} numerically\footnote{It is convenient  to solve the equations of motion \eqref{eqH}, \eqref{eqphis} by rewriting them using the number of e-folds as independent variable $d{\rm N} = H dt$. } with the values of the parameters shown in Table \ref{Table1}. We  fixed the flux number $q$, while we vary the wrap number $p$. However, this is not the only possibility and there is a wider range of $p,q$ values that can be chosen to obtain successful slow-roll fat inflation with the smallest eigenvalue of the scalar mass squared satisfying $\lambda >H^2$.  
Note that once we fix $(N,g_s,u)$ the string and compactification scales are fixed. For the values in Table  \ref{Table1}, the string scale is $M_s \sim 2\times 10^{-3} M_p$, while the compactification scale is set by 
${\mathcal V}_6^{1/6}\sim 13 \,\ell_s$, which gives, for the parameters in Table \ref{Table1}, $M_c\sim 1.53\times 10^{-4}\Mp$. On the other hand, the scale of inflation turns out to be $H\sim  10^{-5}\Mp$ for the  5 choices of $p$ we take (see Table \ref{Table2}).

\begin{table}[H]
\begin{center}
\centering
\begin{tabular}{| l | c | c | c | c | c | c | c |}
\hline
\cellcolor[gray]{0.9} $ N$ & \cellcolor[gray]{0.9} $g_s$ & \cellcolor[gray]{0.9} $\ell_s$  & \cellcolor[gray]{0.9} $u$ & \cellcolor[gray]{0.9} $q$ & \cellcolor[gray]{0.9} $a_0$ & \cellcolor[gray]{0.9} $a_1$ &\cellcolor[gray]{0.9} $b_1$\\
\hline \hline
 $1000$ & $0.01$  & 501.961 & $50 \ell_s$ & 1 &  $0.001$ & $0.0005$ & $0.001$\\
\hline
\end{tabular}
\end{center} 
\caption {Parameter's values for the slow-roll fat inflation example discussed in the text. Note that $\ell_s$ is given in Planck units.}
\label{tab:1}\label{Table1}
\end{table}

\begin{figure}[H]
  \begin{center}
    \includegraphics[width=0.5\linewidth]{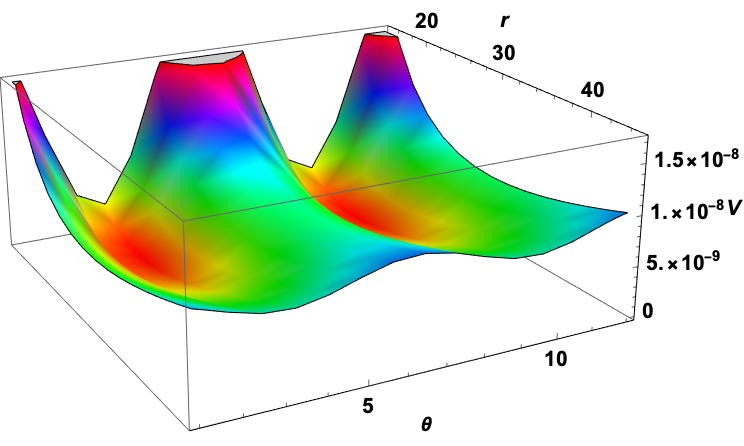}
 \end{center}
\caption{The scalar potential for the parameter values in Table \eqref{Table1}. The value of the minimum does not change when we change $p$. The minimum of the potential is located at $r_{min}=21.414$, $\theta = (2n+1)\pi$, $n\in {\mathbb Z}$. 
The potential and $r$ coordinate are given in Planck units. 
 }\label{potLT}
\end{figure}

Although both fields are evolving and thus a  decay constant for the angular variable cannot be defined, we can define an {\em instantaneous  decay constant} $f$ by  
\be\label{dc}
f= \sqrt{g_{\theta\theta}}.
\ee
It remains approximately constant  during the  first 60-50 efolds (before the end) of inflation with $f_{60}/f_{50}\sim 0.9902$ and grows to about $f_{60}/f_{end}\sim 0.8665$  by the end of inflation. 
In Table \ref{Table2} we give the values of the (average value between ${\rm N}=(60-50)$) instantaneous decay constant for five different choices of $p$ for the parameters' choice in Table \ref{Table1}. We also give the initial conditions for the angular and radial fields as well as the total number of e-folds achieved. 
 In Figure \ref{potLT} we show the  potential in Planck units for the parameter values in Table \ref{Table1}. The minima are  located at $(r_{min}, \theta_{min}) = (21.414,(2n+1)\pi)$, $n\in {\mathbb Z}$ and are  independent of the wrapping number $p$. The  minima of the potential are positive and thus we use $V_0$ to shift this dS minimum to Minkowski as discussed before.

\begin{table}[H] 
\begin{center}
\centering
\begin{tabular}{| l | c | c | c | c | c | c |}
\hline
\cellcolor[gray]{0.9} $p$ & \cellcolor[gray]{0.9} $f/\Mp$  & \cellcolor[gray]{0.9} $\theta_{initial}$ & \cellcolor[gray]{0.9}  ${\rm N}_{tot}$   \\
\hline \hline
$7$ & $ 7.49$  &   1.15 &  $90.79$   \\
\hline
$6$ & $6.89$   &   1.10 &  $83.19$  \\
\hline
$5$ & $6.22$   &   0.95 &  $83.47$   \\
\hline
$4$ & $5.51$    &  0.76 &  $84.33$  \\
\hline
$3$ & $4.71$    & $0.55$ & $83.05$  \\
\hline
\end{tabular}
\end{center} 
\caption{Instantaneous decay constants \eqref{dc} for different values of the wrapping number $p$ for the case study with $r_{min}=21.414$ and $\theta_{min}= \pi$, using values of the parameters in  Table  \ref{Table1} (here $f$ is the average value between 60-50 e-folds before the end of inflation). The initial conditions used for $\theta$ and total number of e-folds achieved are also given and in all cases $r_{initial}=4$.}
\label{Table2}
\end{table}

\begin{figure}[H]
  \begin{center}
    \includegraphics[width=0.55\linewidth]{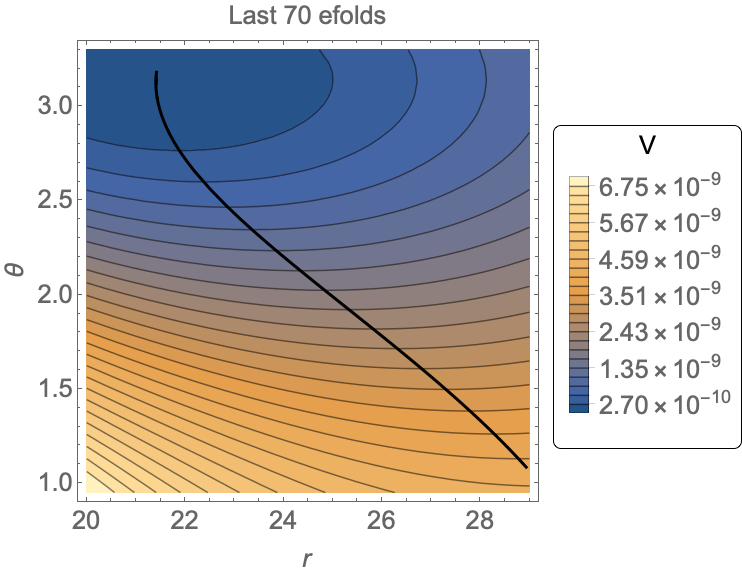} \qquad 
      \includegraphics[width=0.46\linewidth]{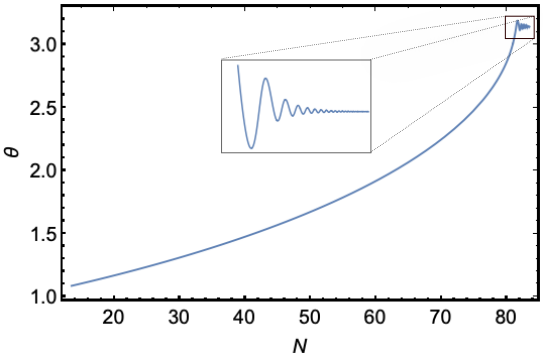} \quad\includegraphics[width=0.46\linewidth]{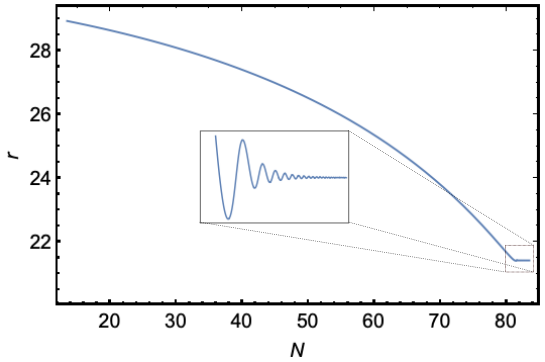}
 \end{center}
\caption{ Fields' trajectory in the potential (upper plot)  and their evolution (lower plots) for the case $f\simeq 6.22$ in Table \ref{Table2}  and parameters given in Table \ref{Table1}.}\label{Trajectories2}
\end{figure}

\begin{figure}[H]
  \begin{center}
    \includegraphics[width=0.54\linewidth]{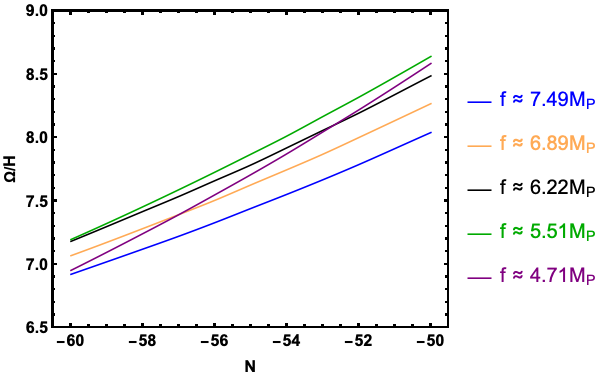}
    \includegraphics[width=0.41\linewidth]{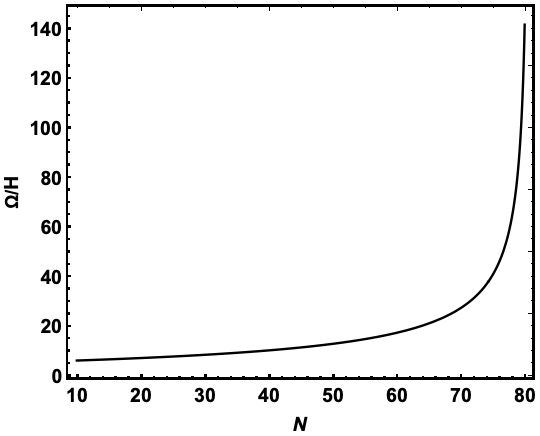} 
 \end{center}
\caption{Turning rate  comparison during the first 10 e-folds for the examples in Table \ref{Table2} (left) and   turning rate for the full inflationary evolution for the case $f\simeq 6.22$ (right). In these examples  $\Omega \sim 10^{-4}\Mp$.}\label{TurnLT}
\end{figure}

\begin{figure}[H]
  \begin{center}
    \includegraphics[width=0.6\linewidth]{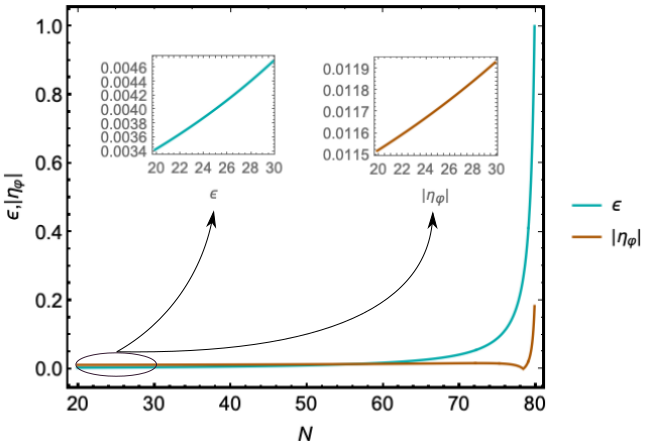}  
 \end{center}
\caption{Slow-roll parameter's evolution for the $f\simeq 6.22$ case. Here  $\eta_{\varphi}\equiv- \frac{\ddot \varphi}{H\dot\varphi}$ (note that $\eta=-2\eta_\varphi+2\epsilon$). }\label{SRparamsLT}
\end{figure}

\begin{figure}[H]
  \begin{center}
    \includegraphics[width=0.43\linewidth]{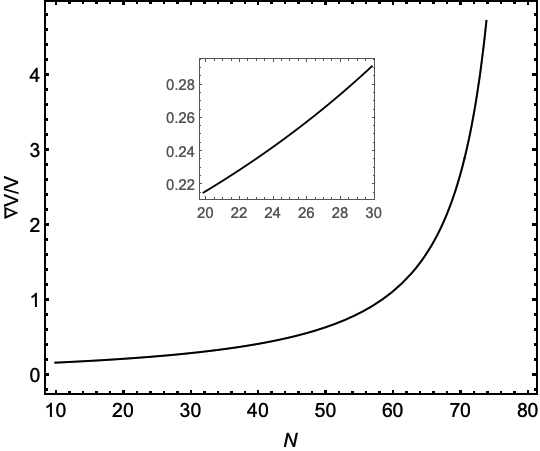} \qquad \includegraphics[width=0.45\linewidth]{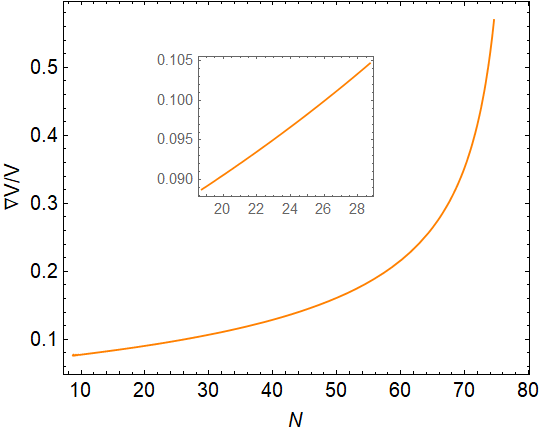}
 \end{center}
\caption{Evolution of $\nabla V/V$ as defined in the swampland constraint \eqref{swamp1} discussed in section \ref{sec:swamp} for the fat inflation case (left) with $f\simeq 6.22$ given in Table \ref{Table2}  and parameters given in Table \ref{Table1}; and small turn case (right) for $f\simeq 6.54$ in Table \ref{Table4} and parameters in Table \ref{Table3}.}\label{Fig_c}
\end{figure}

In figure  \ref{Trajectories2} we show the scalar fields' trajectories along the full inflationary evolution for the case with $f\simeq 6.22$ and other parameter values in Tables \ref{Table1} and \ref{Table2}.  The radial field quickly settles to its displaced minimum at $V(\theta_{initial}, r_{disp})$ and follows it throughout the evolution, as the angular coordinate evolves. Both fields eventually reach their minimum and start oscillating around it.  
For all values of $p$, the turning rate $\Omega/H>1$ as shown in figure \ref{TurnLT}. In all the examples, the dimensionful turn  is small and of order $\Omega\sim 10^{-4}\Mp$. The Hubble parameter on the other hand is of order $H\sim10^{-5}\Mp$ as expected for natural inflation. As we discussed above, the minimum eigenvalue of the mass matrix is larger than the Hubble scale and for all examples it is $\lambda/H\sim 10$. The slow-roll parameters are shown in figure \ref{SRparamsLT} for the $f\simeq 6.22$  example. 
We finally show in figure \ref{Fig_c} the value of $\nabla V/V$ (\eqref{swamp1}), relevant for the swampland constraints (see discussion in section \ref{sec:swamp}), which starts  from  around $0.22$ at ${\rm N_*\sim 60}$ and grows to about $\sim 10^2$ at the end of inflation (right plot).
We have therefore an example of a slow-roll inflationary evolution with large turning rate and only heavy scalar fields.  It is easy to check that  the values of $(p,q)$  are consistent with the backreaction constraints discussed in section \ref{sec:back}. 
%

\subsubsection*{Cosmological parameters}

We now discuss the inflationary predictions for the primordial spectra in the D5-brane fat inflationary model. The dynamics of the linear perturbations  is described  by  equations \eqref{QT}, \eqref{QN}, while the masses of the adiabatic  and entropy modes are given by \eqref{amass} and \eqref{emass}. 
For the model discussed in this section, the scalar manifold curvature ${\mathbb R }$ is negative and large ${\mathbb R } \sim - 3\times 10^4 \Mp^{-2}$ during inflation, however, it does not trigger any instability. Indeed, the mass of the entropy mode  is much larger than $H$, $M/H\sim 10^3$ and thus it decays and can be integrated out. Moreover, $M_{eff}/M\sim 1$ and therefore, the speed of sound is essentially one (see eq.~\eqref{cs}). 
We have also checked the adiabaticity condition \cite{CAP} 
\be\label{Adia}
{\mathcal A}=\left|\frac{\dot\Omega }{M\Omega}\right| \ll 1\,,
 \ee
holds in our case with ${\mathcal A}\sim 5 \times 10^{-4}$.

\begin{figure}[H]
  \begin{center}
    \includegraphics[width=0.85\linewidth]{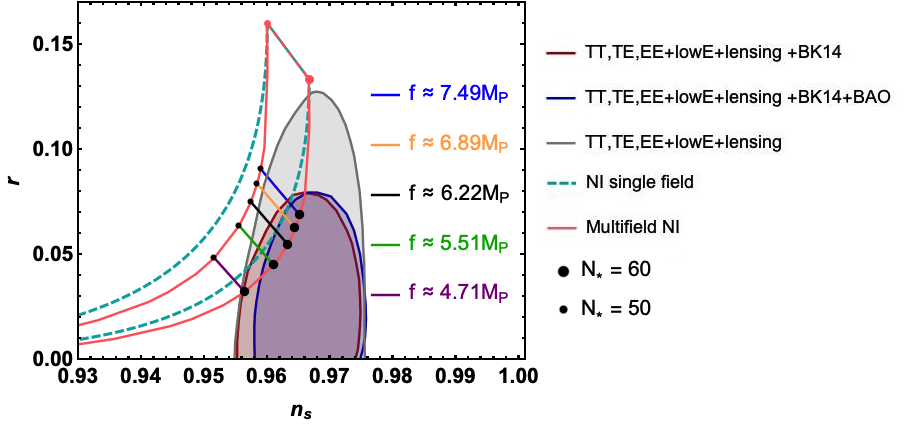}
 \end{center}
\caption{The $(n_s,r)$ plane for the D5-brane multifield fat inflation model discussed in the text with parameters given in Tables \ref{Table1} and \ref{Table2}. The shaded regions are the Planck 95\%CL regions as indicated. The single field natural inflation predictions are indicated by the cyan dashed curve, while the fat D5-brane predictions follow the continuous curve. The effect of the heavy inflatons and large turns move the predictions to the best fit region, even with $c_s\simeq1$ (see main text).}\label{nsr}
\end{figure}

We therefore  use  the standard formulae  for the cosmological parameters in terms of the slow-roll  parameters \eqref{epsilonH}, \eqref{etaH}.  In terms of these the spectral index and the tensor to scalar ratio are given by\footnote{The running of the spectral index turns out to be very small $\alpha_s\sim 10^{-4}$. }
\be
n_s= 1- 2\,\epsilon - \eta \,, \qquad r= 16 \epsilon \,,
\ee
with the latest Planck data \cite{Planck18} giving the values  
\bea
&&n_s= 0.9649 \pm 0.0042   \quad ({\text{at  68\%  CL}}) \,,  \\
&&r <0.10    \quad ({\text{at  95\%  CL}})\,,
\label{nsr_bound}
\eea
while the BICEP2/Keck  BK14 combined analysis gives $ r < 0.064$. 
The amplitude of the power spectra are given by
\be
\Delta_s^2 =\frac{1}{8\pi^2\Mp^2} \frac{H^2}{\epsilon} \sim  2.1\times 10^{-9} \,, \qquad \Delta_t^2 =\frac{2 H^2}{\pi^2\Mp^2}  \,.
\ee
Here all quantities are evaluated at horizon crossing at about $60-50$  efolds before the end of inflation. 
Our final choice of parameters in Tables \ref{Table1} and \ref{Table2} is such that the amplitude of the power spectra match observations. 
In figure \ref{nsr} we show the $(n_s,r)$ plane for the D5-brane multifield fat inflation model discussed above with parameters given in Tables \ref{Table1} and \ref{Table2}.  The single field natural inflation predictions are indicated by the cyan dashed curve, while the multifield D5-brane predictions follow the continuous curve. The effect of the heavy inflatons and large turns move the predictions to the $95\%$CL region, even when $c_s\simeq1$. This can be understood as the slope of the potential and thus the inflationary trajectory changes when the masses of the scalar fields increase. Therefore, the velocities and accelerations will change, giving slightly different values of the slow-roll parameters and thus  cosmological observables\footnote{We haven't added the uncertainty in the number of efolds between horizon crossing and the end of inflation, ${\rm N}_*$, coming from reheating after inflation. In the case when the post inflationary evolution is dominated by scalars, it is possible that $ {\rm N}_*$ is shifted to larger values, providing a better fit for natural inflation \cite{MZ}.}.  
Although it is interesting that fat inflation gives different predictions to single field, the $(n_s,r)$ plane is not enough to distinguish between them. We therefore give a first look into the non-Gaussianity following \cite{Evan}.

\bigskip

\subsubsection*{Primordial Non-Gaussianity $f_{\mathrm{NL}}^{local}$}

We now compute the local type non-Gaussianity,
$f_{\mathrm{NL}}^{local}$, associated with the previous fat
inflationary trajectories
\footnote{See \cite{Gong:2013sma} where an analytical expression for the equilateral non-Gaussianity  for  a simple two-field inflation model is presented.}. We follow the covariant $\delta
{\mathrm{N}}$ formalism of \cite{Evan} for inflationary models on a
curved manifold, where the non-linear parameter
$f_{\mathrm{NL}}^{local}$ takes the standard form
\be\label{fnl}
f_{\mathrm{NL}}=-\frac{5}{6}\frac{\mathrm{N}^{,i}{\mathrm{N}}^{,j}
{\mathrm{N}}_{;ij}}{({\mathrm{N}}_{,k}{\mathrm{N}}^{,k})^2} \,,
\ee
where $i$ refers to $\{r,\theta\}$, comma and semicolon denote the
partial and covariant derivatives with respect to the scalar fields
$\{r,\theta\}$ and the scalar-field metric $g_{ij}$. Notice that we
have removed the label $local$ for convenience.

In order to calculate $f_{NL}$ numerically, we use the method of
finite differences for the derivatives (e.g. ${\rm N}_{,r}=({\rm
N}(r+\Delta r,\theta)-{\rm N}(r-\Delta r,\theta))/(2\Delta r)$, etc.),
and integrate ${\rm N}(r,\theta)$ from the horizon crossing of the
relevant mode, ${\rm N}_{*}$, to the end of inflation, $N_{end}$,
defined where ${\epsilon=1}$. We choose modes in the range from ${\rm
N}_{*}=50$ to ${\rm N}_{*}=70$ prior to the end of inflation. Given
that the final result for $f_{NL}$ is very sensitive to tiny values of
$(\Delta r,\Delta\theta)$ at horizon crossing, we average over a few
possible larger values ($\Delta r\approx\mathcal{O}(10^{-1})$ and
$\Delta\theta\approx\mathcal{O}(10^{-3})$), making sure that their
dispersion is roughly two orders of magnitude smaller than the
resulting $f_{NL}$. Moreover, we have also checked that slight
differences in the definition of the end of inflation do not change
the final value of $f_{NL}$.

\begin{figure}[H]
  \begin{center}
    \includegraphics[width=0.39\linewidth]{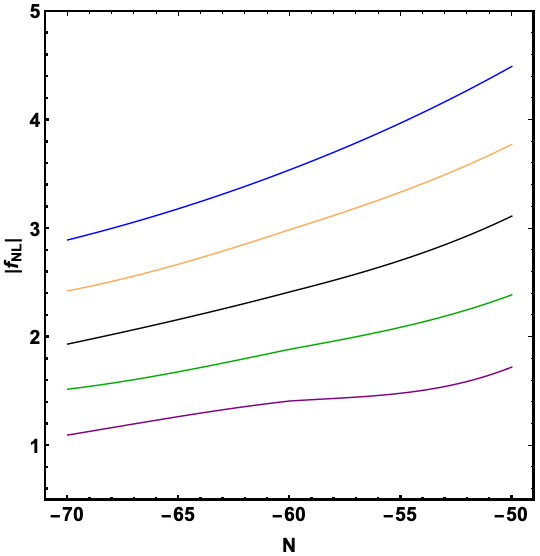}
    \includegraphics[width=0.56\linewidth]{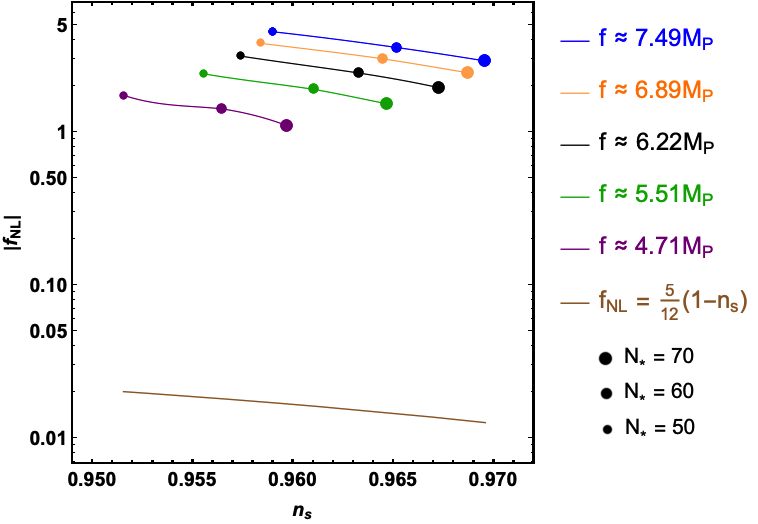}
 \end{center}
\caption{(Left) $|f_{NL}|$ as a function of the number of e-folds {\rm
N} for the five different cases of decay constants and initial
conditions presented in Table \ref{Table2}, with error bars (as the standard
deviation of the averaged value for $f_{NL}$ using 9 different
combinations of $(\Delta r,\Delta\theta)$) of the order of $10^{-2}$.
(Right) $|f_{NL}|$ vs $n_s$ for the same decay constants. All values
of $f_{NL}$ for these fat inflation realisations are negative, and
deviate from from the single field consistency condition (brown solid
line). }\label{NGLT}
\end{figure}

In figure \ref{NGLT} we show the results for $f_{NL}$ for the five
decay constants discussed ealier, and find that they are all negative
and of order $\mathcal{O}(1)$, falling within the most recent bounds
by Planck, $f_{\mathrm{ NL}}^{local} = -0.9 \pm 5.1$
\cite{non-gaussian}. Furthermore, once comparing our $f_{NL}$ results
with the single clock consistency relation
$f_{NL}=\frac{5}{12}(1-n_s)$ \cite{Malda},
they clearly depart from the single field model (see the right plot in
figure \ref{NGLT}).

\subsection{D5-brane inflation with a light field: small turns}\label{ST}

We now present an example of a choice of parameters where the  turning rate is smaller than one  and one of the field's is lighter that the Hubble parameter. That is, a ``standard"  hierarchy for the mass of the fields holds: $M_1 \lesssim H < M_2 $. 
In particular, $M_1/H\sim 0.35$ at ${\rm N_*}=60-50$ and in this case, $\nabla V/V$  is slightly smaller than in the fat inflation example above with  $\nabla V/V\gtrsim 0.1$ at at ${\rm N_*\sim 60}$ for the $f\sim 6.54$ case (see left plot in figure \ref{Fig_c}). 
 This example illustrates the differences between the two types of inflationary evolution that can arise in multifield models. 

\begin{figure}[H]
  \begin{center}
    \includegraphics[width=0.55\linewidth]{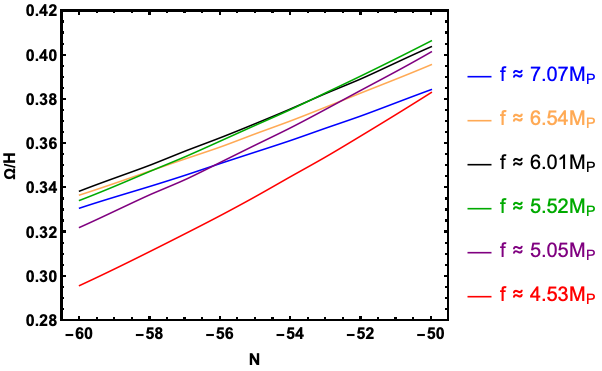}
     \includegraphics[width=0.4\linewidth]{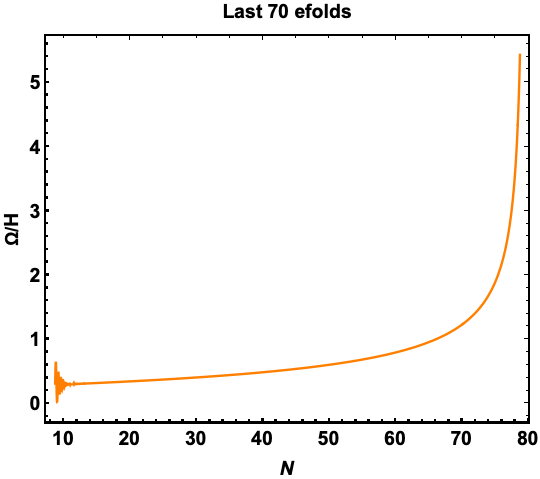}
 \end{center}
\caption{Turning rate  comparison during the first 10 e-folds for the examples in Table \ref{Table4} (left) and   turning rate for the full inflationary evolution for the case $f\simeq 6.54$ (right). In these examples  $\Omega \sim 10^{-5}\Mp$.}\label{TurnST}
\end{figure}

\begin{figure}[H]
  \begin{center}
    \includegraphics[width=0.8\linewidth]{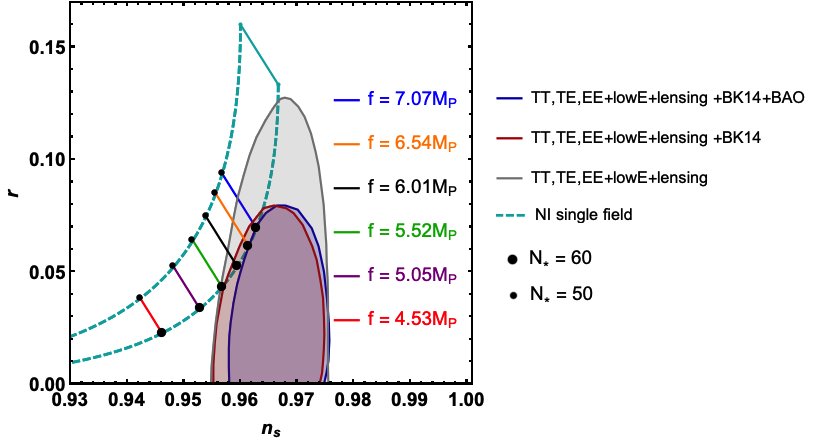}
 \end{center}
\caption{The $(n_s,r)$ plane for the D5-brane multifield  inflation model with small turning rate with parameters given in Tables \ref{Table3} and \ref{Table4}. The shaded regions are the Planck 95\%CL regions as indicated. The predictions fall exactly along the single field natural inflation curve  (cyan dashed curve).}\label{STnsr}
\end{figure}

The parameters' values are shown in Table \ref{Table3}.  Instantaneous decay constant,  initial conditions, and total number of e-folds achieved are given in Table \ref{Table4}.  The instantaneous decay constant in this case remains almost unchanged during the whole inflationary evolution with $f_{60}/f_{end}\sim 0.9998$. In figure \ref{TurnST} we show the turning rates for this set of parameters and in figure \ref{STnsr} we show the predictions for the spectral tilt and the tensor-to-scalar ratio. As it is clear from the plot, the multifield D5-brane inflation is indistinguishable from single field natural inflation at linear order in perturbations.
In this example too the mass of the adiabatic mode is small w.r.t.~$H$, while $M\gg H$ and $M_{eff}\sim M$, so that $c_s\sim 1$. Finally, the adiabaticity condition \eqref{Adia}
 in this case gives ${\mathcal A}\sim 10^{-3}$. 
 
\begin{table}[H]
\begin{center}
\centering
\begin{tabular}{| l | c | c | c | c | c | c | c |}
\hline
\cellcolor[gray]{0.9} $N$ & \cellcolor[gray]{0.9} $g_s$ & \cellcolor[gray]{0.9} $q$ & \cellcolor[gray]{0.9} $u$ & \cellcolor[gray]{0.9} $\ell_s$ & \cellcolor[gray]{0.9}$a_0$ & \cellcolor[gray]{0.9}$a_1$ & \cellcolor[gray]{0.9} $b_1$\\
\hline \hline
 $1000$ & $0.01$  & 70 & $50 l_s$ & 501.961 &  $0.1$ & $0.0001$ & $0.0001$\\
\hline
\end{tabular}
\end{center} 
\caption {Parameter's values. Note that $\ell_s$ is given in Planck units. That is, $M_s = 2\times 10^{-3} \Mp$.}
\label{Table3}
\end{table}

\begin{table}[H] 
\begin{center}
\centering
\begin{tabular}{| l | c | c | c | c | c | c |}
\hline
\cellcolor[gray]{0.9} $p$ & \cellcolor[gray]{0.9} $f/\Mp$  & \cellcolor[gray]{0.9} $r_{initial}$ & \cellcolor[gray]{0.9} $\theta_{initial}$ & \cellcolor[gray]{0.9}  ${\rm N}_{tot}$   \\
\hline \hline
 $90$ & $7.07$   & $400$ &  105.873 &  $81.83$   \\
\hline
$77$ & $6.54$  & $410$ &  105.973 &  $79.24$   \\
\hline
$65$ & $6.01$   & $430$ &  106.073 &  $75.62$  \\
\hline
$55$ & $5.52$   & $500$ &  106.173 &  $73.06$   \\
\hline
$46$ & $5.05$    & $380$ & 106.373 &  $80.70$  \\
\hline
$37$ & $4.53$    & $410$  & $106.473$ & $75.64$  \\
\hline
\end{tabular}
\end{center} 
\caption{Decay constants for different values of the wrapping number $p$ for the case study with $r_{min}=456.797$ and $\theta_{min}=33 \pi$, using values of the parameters in  Table  \ref{Table1}. The initial conditions used for $(r, \theta)$ and total number of e-folds achieved are also given.}
\label{Table4}
\end{table}    

\begin{figure}[H]
  \begin{center}
    \includegraphics[width=0.35\linewidth]{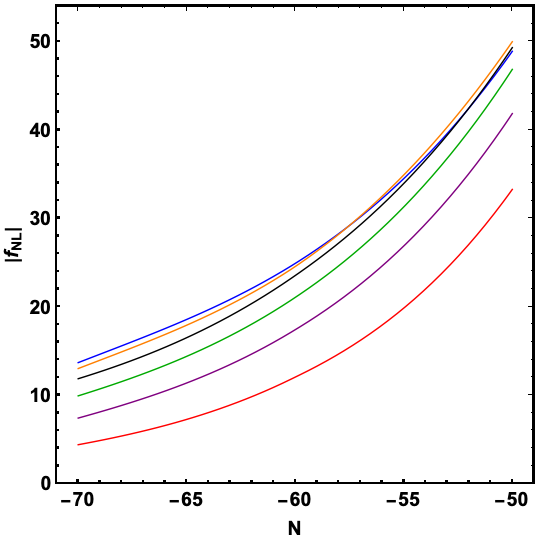}
 \includegraphics[width=0.5\linewidth]{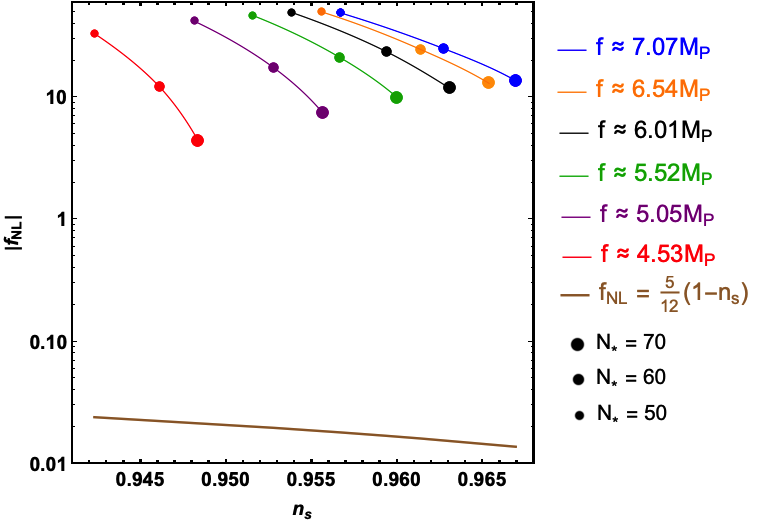}
 \end{center}
\caption{Left image shows the value of $|f_{NL}|$ vs the number of e-folds {\rm N} for the five different cases of decay constants and initial conditions presented in \ref{Table4}. The right image shows  $|f_{NL}|$ vs $n_s$ for the same decay constants. All values
of $f_{NL}$ for these model  are negative. }\label{NGST}
\end{figure}

We finally compute  the non-Gaussianity parameter for this example following the same procedure as before. The results are  shown in figure \ref{NGST} (the value of $f_{NL}$ we find is negative also in this case). In this case, as it is clear from the plot, although the predictions for $(n_s,r)$ are indistinguishable from single filed, the non-Gaussianity parameter is large and falls outside the most recent constraints from Planck. It is interesting that for smaller turns, the non-linear parameter turns out to be much larger. We do not have an intuition for this result and would be interesting to explore this further. Let us note only that  in  \cite{Evan}, it was  found that very different values for $f_{NL}$ are obtained as the trajectory of the inflatons changes.

\section{Discussion}\label{disc}

We have shown that a successful period of slow-roll inflation can be achieved in multifield scenarios even when the masses of the scalar fields are heavier than the Hubble scale, that is, $H < M_{inf}$, where $M_{inf}$  is the mass of the ``lightest"  field. 
We call this attractor {\em fat inflation} to stress that it is the masses of all the scalar fields, which are heavier than the Hubble scale, rather than the masses  of the quantum fluctuations. Indeed, in terms of the masses of the fluctuations, the mass of the adiabatic mode is given in terms of the slow-roll parameters, and therefore it is always  smaller than $H$ during slow-roll, while the isocurvature mass(es) might be heavy or light, with different cosmological implications \cite{Anacs,Anacs1,CP,Garcia-Saenz:2018ifx,Orbital}. 

This is a non-trivial result, as it is commonly believed that large contributions to the masses of the inflatons might spoil slow-roll inflation, a phenomenon that  goes under the name of  $\eta$-problem. However, we have seen that large contributions to the masses do not necessarily spoil multifield slow-roll inflation. We showed that this scenario unavoidably has large turning rates $\Omega/H$, and therefore non-geodesic trajectories. Fat inflation thus evades the $\eta$-problem with large turns in multifield scenarios. 
Fat inflation opens up a new possibility for multifield inflation in which large turns and thus non-geodesic motion are unavoidable, with interesting implications for the dS swampland conjectures and possible cosmological implications that may be testable in the forthcoming years. As we discussed in the explicit D5-brane example (sections \ref{sec:2}, \ref{Sec:3}), the cosmological predictions differ from single field and may be distinguishable from it via non-Gaussianities. 

In   appendix \ref{Models} we collected  examples of field theory multifield  models in the literature, which happen to belong to the fat slow-roll attractor. These include a recently discussed  three field  model in \cite{Paban} (APR) where the lightest field is sixty times the Hubble scale, $m_1/H\gtrsim 60$, while the heaviest is thousand times heavier $m_3/H\gtrsim 4500$. The sidetrack models, where there is a transition from a standard slow-roll trajectory with a light and a heavy field, to a  fat slow-roll trajectory, with both scalar fields having larger masses than the Hubble scale. 

In sections  \ref{sec:2}, \ref{Sec:3},  we  presented an explicit  example of fat  inflation using a probe D5-brane moving in the warped resolved conifold of a type IIB flux compactification.  
The fat inflatons correspond to the scalar fields associated to the  radial and one  angular directions.  The brane is  assumed to be fixed along the other two angular directions and we assumed also that the closed string moduli can be stabilised using a combination of fluxes and non-perturbative terms. 
The scalar potential for the two-fields has a cosine dependence on the angular direction,  which can be used to realise natural inflation \cite{KT}.
We defined an  instantaneous, field dependent decay constant as $f=\sqrt{g_{\theta\theta}(r)}$, which took superplanckian values realising  a fat natural inflation model. The cosmological parameters differ slightly from single field natural inflation as we showed in figure \ref{nsr}. As we discussed, the speed of sound remains basically one and the difference in the predictions w.r.t.~to single field can be understood by the different behaviour of the slow-roll parameters (or the potential) along the inflationary trajectory when fat fields drive inflation. For comparison, we also presented an example of a set of parameters which gives a standard hierarchy of masses in \ref{ST}. In this case, the predictions coincide with the single field case as shown in figure \ref{STnsr} and thus would be impossible to distinguish between the two cases using only $(n_s,r)$. In both examples, fat and standard inflation, the inflationary trajectory deviates  from a geodesic, which is measured  by the turning rate $\Omega/H$ (see section \ref{sec:1}) which is order one for the standard case and order ten in the fat case (see figures \ref{TurnLT}, \ref{TurnST}). The scalar curvature is negative and large in the fat and standard examples (${\mathbb R}\sim-10^{4}\Mp^{-2}$, ${\mathbb R}\sim-10^{2}\Mp^{-2}$ respectively). However no geometric destabilisation is triggered. In both examples too, the mass of the entropic mode is well above the Hubble scale. 

 We have used the results in \cite{Evan} to compute the local non-Gaussianity, which would be a useful tool to distinguish multifield model predictions from the single field case. For the fat inflationary case, we found that the non-Gaussianity is of order one (see fig.~\ref{NGLT}) and can therefore constitute a powerful tool to distinguish this model from  single field, which predicts a negligible level of non-Gaussianity. The standard example with small turning rate on the other hand gives a much larger value for the $f_{NL}$ parameter (see fig.~\ref{NGST}) and would be ruled out by current bounds. Although we do not have a clear intuition for this result, it has been shown in \cite{Evan} how different trajectories can give completely different values for the non-Gaussian parameter. Although the inflation model studied in \cite{Evan} has tiny turning rates (${\mathcal O}(10^{-3}-10^{-4})$), it holds that also in that case, for the trajectory with larger value of $\Omega/H$, the non-Gaussian parameter is smaller and viceversa. It would be interesting to study this behaviour in more detail, as it could be important to distinguish among  single and multifield models of inflation. 
  
Let us finally comment on the challenges of the D5-brane model. As we have discussed, 
the instantaneous superplanckian decay constant is consistent with the weak coupling $g_s<1$ limit with   a hierarchy of scales given by $\Mp\gtrsim M_s\gtrsim M_c  > H$ ($M_s\sim 10^{-3}$, $M_c\sim 10^{-4}$, $H\sim 10^{-5}$). 
Consistency with the WGC could be understood along the lines discussed in \cite{KPZ}. However, as we mentioned before, the WGC  for axions in its original form can not be straightforwardly applied to the case we consider here, that is, a field dependent decay ``constant" and a non-geodesic trajectory. We leave for future work a careful study of this situation. 
Let us further stress  that we haven't considered a full embedding of the WRC into a controlled type IIB flux compactification. Therefore, we can only estimate the value of $M_c$ and thus $M_s$ from \eqref{eq:Mp} and our choice for $r_{UV}=u$. We have thus simply assumed that the closed string moduli are heavier than the open string moduli, the brane positions which drive inflation. 
On the other hand, it turns out that the heaviest eigenvalue of the mass matrix for the D5-brane model is of the order of the string scale in both examples and therefore  heavier than the closed string moduli, assumed to be fixed. This is a drawback of our model in its present form and it would be necessary to make a full embedding of the WRC into a  flux compactification to properly address this problem. Though our D5-brane model is far from being complete, we took the first step towards understanding large turning rate inflation in string theory. 

More generally, in view of our present results, it would be interesting to revisit D-brane models, such as D3-brane multifield inflation, which have  been studied in the standard inflationary attractor with small turns. This will also be important in view of the recent theoretical constraints on standard slow-roll  inflation and forthcoming experiments.

\acknowledgments
We thank Ana Ach\'ucarro, Nana Cabo-Bizet, Carlos N\'u\~nez, Susha Parameswaran, Evangelos Sfakianakis, Gianmassimo Tasinato and Yvette Welling for discussions.   DCh  would also like to thank Armando A.~Roque Estrada for useful discussions on the numerics used in this paper.  IZ thanks Department of Physics, University of Guanajuato, Mexico for hospitality during the early stages of this work. DCh, OLB and IZ thank also Casa Matem\'atica Oaxaca (CMO-BIRS) for hospitality and excellent facilities while part of this work was done.
 This work was partially supported by a Mexican Academy of Sciences mobility grant (Newton Fund AMC-CONACyT).  OLB and GN were partially supported by CONACyT grants 258982 CB-2015-01 and 286897, and by DAIP-UG (2018 and 2019). GN also acknowledges the support of the Instituto Avanzado de Cosmolog\'ia A.C.~and the DataLab DCI-UG. DCh is supported by a CONACyT scholarship and was partially supported by DAIP-UG grant. RCh is supported by and STFC postgraduate scholarship in data intensive science.   IZ is partially supported by STFC, grant ST/P00055X/1. 


\begin{appendix}

\section{Field theory models of fat inflation}\label{Models}

In this appendix we first describe how to construct  the simplest field theory  model for two fields leading to  the fat inflation attractor with large turning rate. We then 
collect some field theory multifield inflation examples  in the literature that happen to be {\em fat field inflation} models  and compare them with some ``light field"  (that is where $M_{inf}<H$) examples also in the literature. 

As we discussed in section \ref{SR}, when the minimal eigenvalue of the mass matrix is larger than the Hubble parameter, the multifield slow-roll condition is satisfied and the turning rate is large, indicating a non-geodesic trajectory (see section \ref{SR}). It is thus clear that some interaction between the scalar fields is necessary, which can either come from the kinetic terms or the scalar potential. The simplest possibility is  to consider a flat scalar manifold in polar coordinates, that is,  $g_{ab} = {\rm diag}(1,\rho^2)$ for the scalar fields $\rho, \theta$ with potential $V(\rho,\theta)$. The eigenvalues of the mass matrix in this case take the simple form:
\be
\lambda_{\pm} = \frac{1}{2}\left(V_{\rho\rho} + \frac{V_\rho}{\rho} +\frac{V_{\theta\theta}}{\rho^2} \pm 
\sqrt{\left(V_{\rho\rho} - \frac{V_\rho}{\rho} -\frac{V_{\theta\theta}}{\rho^2}\right)^2 + \frac{4}{\rho^2}\left(V_{\theta\rho}-\frac{V_{\theta}}{\rho}  \right)^2}\right)\,,
\ee
where $V_{a}, V_{ab}$ denote partial derivatives of $V$ with respect to the scalar fields. 
If we now consider a  scalar potential  of the form $V= \frac{1}{2}M^2\rho^2 + W(\theta)$, the eigenvalues simplify to
\be
\lambda_{\pm} = M^2 +\frac{W_{\theta\theta} \pm \sqrt{W_{\theta\theta}^2 + 4 W_{\theta}^2}}{2\rho^2} \,.
\ee
As we mentioned, the fat inflationary attractor occurs when $\lambda_- >H^2\sim V = \frac{1}{2}M^2\rho^2 + W(\theta)$. Therefore, for a sufficiently large value of $M$  and when $\rho$ is close to its minimum, the fat inflation condition can be satisfied. Moreover, the potential for $\theta$, $W(\theta)$, is taken such that successful inflation does indeed occur. The simplest possibility is to consider $W(\theta) = m^2 \theta^2/2$ with $M\gg m$. Other possibilities are $W(\theta) = \Lambda^4 \theta^4/4$ or $W(\theta) =\Lambda^4( 1+ \cos(m \theta))$ with $M \gg \Lambda$. This last example is listed below in table \ref{tab:5} (AAW2).

We now collect in table \ref{tab:5}  some examples of field theory multifield inflation models in the literature that happen to be fat inflation models. We list the model's name,  $\Omega/H$, the  mass hierarchy and scalar curvature ${\mathbb R}$. (In these models $\Omega<\Mp$). We include also a multifield supergravity ``light field" inflationary example, in which the fields follow an almost geodesic trajectory, that is, where $\Omega/H\ll1$. 

The first models  in \ref{tab:5}, \textit{Orbital inflation} \cite{Orbital}, \textit{Spiral inflation} \cite{Spiral}  and \textit{Racetrack inflation} \cite{Racetrack} together with AAW2 \cite{AAW}, have all the usual mass hierarchy\footnote{Here we use   the Lagrangean presented  in \cite{AAW}, which is a two field model with a flat scalar manifold (${\mathbb R=0}$) written in polar coordinates with a potential $V=V_0\left[M^2/2(\rho-\rho_0) + (1+\cos{(m \theta)})\right]$. For the values we give in tables \ref{tab:5}, \ref{tab:6}, we use the following parameters $(m=0.002, \rho_0=0.0001)$ and $ M=100$ for  AAW1 while $M=0.15$ for AAW2 in Planck units ($V_0$ can then be adjusted to match the amplitude of the power spectrum).}. 
Compared to the other models,  racetrack inflation has very small turning rate: $\Omega /H \sim O(10^{-4})$ and thus follows an almost geodesic trajectory (see section \ref{SR}). 
 AAW2, on the other hand is characterised by $\Omega/H \gtrsim 2$; this is possible when $V_{TT}/V >1$ even when $\lambda <0$ and thus smaller than $H^2$ (see section \ref{SR}). 

As {\em fat inflation} models with \textit{large} turning rates, we show an example of  two-field natural inflation model discussed in \cite{AAW} (AAW1), the recent three field model in \cite{Paban} (APR) and  the sidetrack models in \cite{Garcia-Saenz:2018ifx}. 
These all have  large  $\Omega/H$, and only the sidetrack models have a non-zero negative curvature $\mathbb{R}$. 
In table \ref{tab:6} we show  the ratio between the masses and the Hubble parameter for AW1, APR and the sidetrack models (both the minimal an hyperbolic examples have similar mass hierarchies).

\begin{figure}[H]
  \begin{center}
  \includegraphics[width=0.47\linewidth]{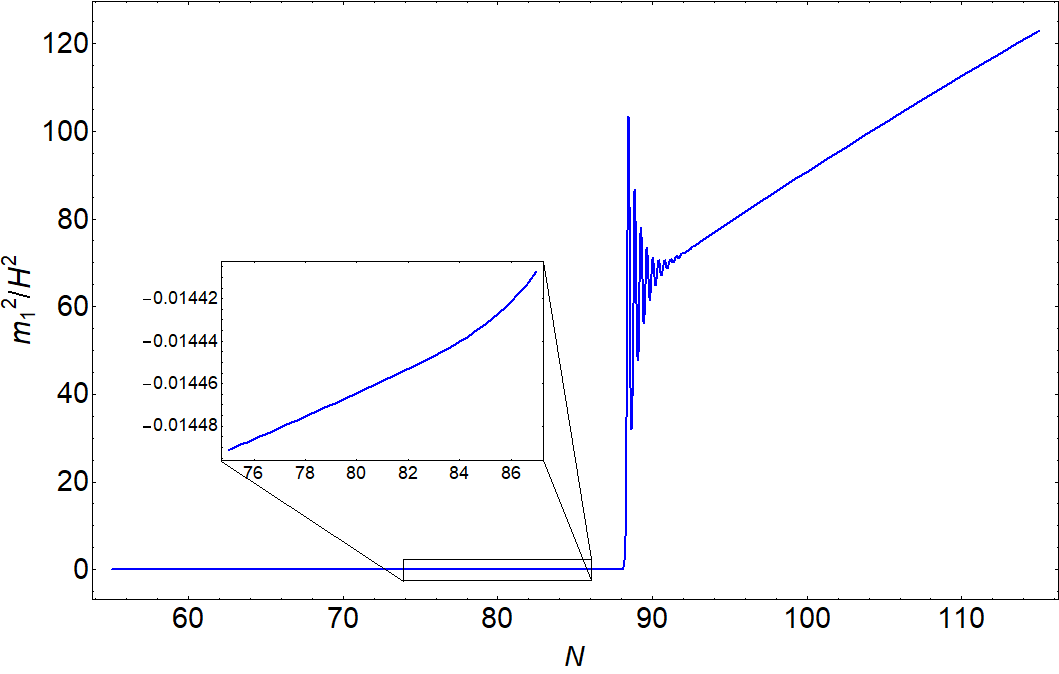} \quad\includegraphics[width=0.47\linewidth]{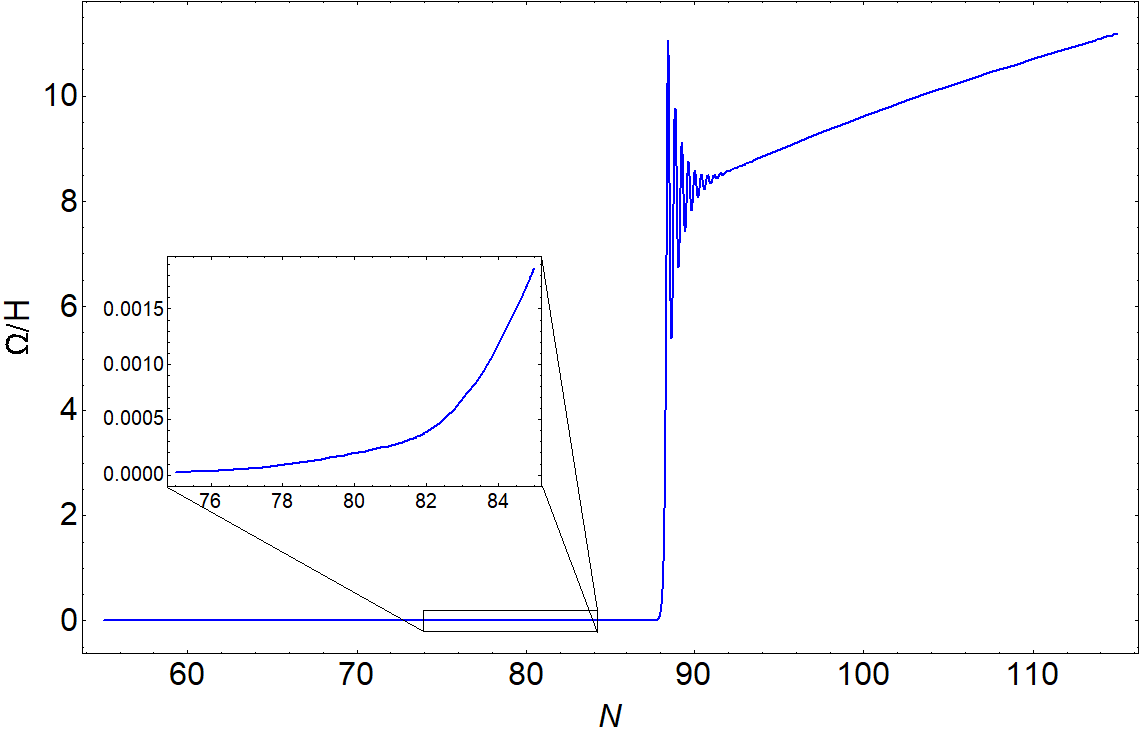}
 \end{center}
\caption{Comparison of the mass of the lightest scalar field and the Hubble parameter (left) and turning rate $\Omega/H$ (right) for the minimal sidetrack (NI) model during the first part of  inflation.  }\label{fig_Sidetrack}
\end{figure}

\begin{table}[htbp!]\small
\begin{center}
\begin{tabular}{| l | c | c | c | c | c | c |}
\hline
\cellcolor[gray]{0.9} {\bf Model} &  \cellcolor[gray]{0.9}$\Omega/H$ & \cellcolor[gray]{0.9} {\bf mass spectrum} & \cellcolor[gray]{0.9}$\mathbb{R}$($\Mp^{-2}$)  \\
\hline \hline
Orbital Inflation \cite{Orbital}&  $\sim -0.2 $  &  $ m_1<m_2<H $ & $0$  \\
\hline
Spiral Inflation \cite{Spiral}   & $\sim -0.12 $ &  $m_1< H < m_2$ & $0$  \\
\hline
Racetrack \cite{Racetrack} & $\sim 6 \times 10^{-4}$ &  $m_1< H < m_2$ & $-\frac{2}{3}$  \\
\hline
AAW2 \cite{AAW}  & $\sim 2$ &  $ m_1 < H  < m_2 $ & $0$  \\
\hline
Minimal sidetrack (NI)  \cite{Garcia-Saenz:2018ifx} &$  \sim 70$   &  $H < m_1 < m_2$ & $-\frac{4{\mathcal M}^2}{({\mathcal M}^2 + 2\chi^2)^2} $ \\
\hline
Hyperbolic sidetrack (NI)  \cite{Garcia-Saenz:2018ifx} & $\sim 163$ &  $ H < m_1 < m_2 $ & $-\frac{4}{({\mathcal M})^2}$ \\
\hline
Minimal sidetrack (Staro) \cite{Garcia-Saenz:2018ifx}  &$  \sim 16$   &  $H < m_1 < m_2$ & $-\frac{4{\mathcal M}^2}{({\mathcal M}^2 + 2\chi^2)^2} $ \\
\hline
Hyperbolic sidetrack (Staro) \cite{Garcia-Saenz:2018ifx}  & $\sim 150$ &  $ H < m_1 < m_2 $ & $-\frac{4}{({\mathcal M})^2}$ \\
\hline
AAW1  \cite{AAW} & $\sim 12$ &  $ H < m_1  < m_2 $ & $0$  \\
\hline
APR \cite{Paban}  & $\sim 61$ &  $H <  m_1 < m_2 < m_3$ & $0$ \\
\hline
\end{tabular}
\caption{Inflationary models illustrating fat and light field inflation. Here ${\mathcal M}$ is the curvature scale and $\chi$  is one of the fields. For all the  models (except APR) we give the value of $\Omega/H$ at the start of  the last 60 e-folds before the end of inflation (where $\epsilon\sim 1$), after which these parameters increase similarly to our D5-brane example (see figs.~\ref{TurnLT}, \ref{TurnST}). In the APR model inflation does not end, so the values of the parameters are given at the start of inflation. In this example, $\Omega$ decreases, while $\Omega/H$ remains almost constant (see \cite{Paban}).}
\label{tab:5}
\end{center} 
\end{table}

\begin{table}[htbp!]
\begin{center}
\centering
\begin{tabular}{| l | c | c | c | c | }
\hline
\cellcolor[gray]{0.9}  {\bf Model} & \cellcolor[gray]{0.9}  $m_3/H$ & \cellcolor[gray]{0.9}  $m_2/ H$ & \cellcolor[gray]{0.9}  $m_1 /H $  \\
\hline \hline 
Sidetrack (NI-Staro)  &-- &$\gtrsim 35$ & $ \gtrsim 30 $  \\
\hline
AAW1 & -- & $ \gtrsim 13$ &  $\gtrsim 10 $   \\
\hline
APR & $ \gtrsim 4500$ &  $\gtrsim 632 $ &   $\gtrsim 60$  \\
\hline
\end{tabular}
\end{center} 
\caption {Ratio of masses to the Hubble parameter  for the fat inflationary models as indicated.  Again, we give the value of the masses at the start of  the last 60 e-folds before the end of inflation (and at the start of inflation for APR).}
\label{tab:6}
\end{table}

The AAW1 model has a reduced speed of sound as in this case it holds that $M_{eff}>M$, that is $\Omega> M$ with both $M, M_{eff}\gg H$. It has  a relatively mild hierarchy of masses, comparable with  sidetrack. 
The hierarchy of masses results to be way more  dramatic in the APR three-field model, where it is worth noticing that the potential does not have a minimum and therefore  inflation does not end. Finally in the sidetrack models the scalar curvature is  negative 
$\mathbb{R} <0$ and triggers an instability which sends the light field  inflationary attractor ($M_{inf}<H$) to the heavy field inflationary attractor we introduced in section \ref{sec:1}.  
These models  have a small  $\Omega/H$ during the light field attractor, which becomes large when the fields settle into the fat field attractor (see figure  \ref{fig_Sidetrack}).
At the same time, the mass hierarchy changes from the standard $m_1<H<m_2$ to the fat hierarchy  $H<m_1<m_2$. This is shown in figure \ref{fig_Sidetrack}.

\section{Double D5-brane Inflation }\label{App2}

\begin{figure}[H]
  \begin{center}
    \includegraphics[width=0.5\linewidth]{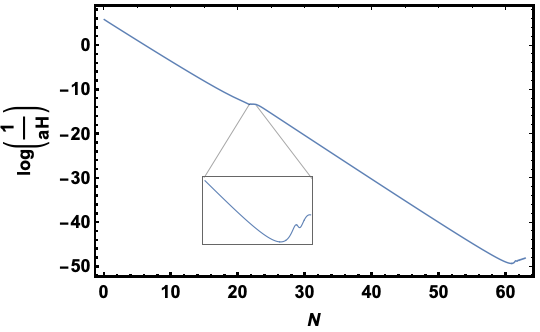}  
    \includegraphics[width=0.45\linewidth]{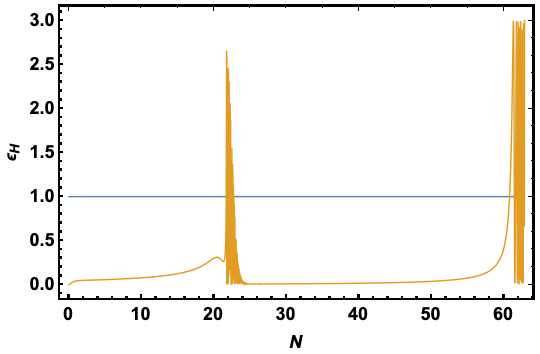}
 \end{center}
\caption{Hubble horizon (left) and  $\epsilon$ (right) evolution for the double inflation example discussed in the text.  }\label{Double2}
\end{figure}

Finally in this appendix we present a possibility for a double inflation realisation using the scalar potential for the D5-brane discussed in the main text. It is clear that this possibility can arise  if one considers initial conditions such that the radial field starts far away enough from its minimum. The behaviour of the potential in this limit is dictated by an $r^4$ power, while in the angular direction it is given by the cosine. In a double inflation realisation, the radial coordinate starts evolving driving a first period of $r^4$ inflation, while the angular coordinate stays frozen until $r$ reaches its shifted minimum, oscillates a  around it for a while  and $\theta$ takes off towards its minimum driving a second period of inflation  driven by the cosine term. Between the two periods of inflation, the slow-roll approximation is broken (see figure  \ref{Double2}): the Hubble horizon starts to increase, and $\epsilon$ becomes larger than one. Interestingly, for this example the value of $\epsilon$ in the first phase of inflation is larger than in the second $\epsilon_2<\epsilon_1$ and thus it can potentially give rise to efficient production of primordial black holes.  
Unfortunately for the $r$ coordinate to drive inflation, it has to start off outside the throat in the WRC, $r_{initial}>r_{UV}$. Moreover, since inflation is driven by a quartic power, the predictions for  CMB scales would lie outside the current observable bounds. It is still interesting to show how such a scenario could arise in a D-brane model.

\begin{table}[H]
\begin{center}
\centering
\begin{tabular}{| l | c | c | c | c | c | c | c | c |}
\hline
\cellcolor[gray]{0.9} $N$ &\cellcolor[gray]{0.9} $g_s$ &  \cellcolor[gray]{0.9} $q$ & \cellcolor[gray]{0.9} $u$ & \cellcolor[gray]{0.9} $\ell_s$ & \cellcolor[gray]{0.9} $a_0$ &\cellcolor[gray]{0.9} $a_1$ & \cellcolor[gray]{0.9}$b_1$\\
\hline \hline
 $1000$ & $0.01$ &  70 & $50 \ell_s$ & 501.96 &  $0.00025$ & $10^{-5}$ & $10^{-5}$\\
\hline
\end{tabular}
\end{center} 
\caption {Parameter's values for the double inflation model discussed in the text. Note that here $\ell_s$ is given in Planck units. }
\label{tab:7}
\end{table}

\begin{table}[H] 
\begin{center}
\centering
\begin{tabular}{| l | c | c | c | c | c | c | c |}
\hline
\cellcolor[gray]{0.9} $q$ &\cellcolor[gray]{0.9} $p$ & \cellcolor[gray]{0.9}$r_{initial}$ & \cellcolor[gray]{0.9}$\theta_{initial}$ & \cellcolor[gray]{0.9} ${\rm N}_{tot}$    \\
\hline \hline
$72$ & $53$ & $149.414$  & $105.773$ & $62.85$  \\
\hline
\end{tabular}
\end{center} 
\caption {Case study with $r_{min}=1.06656$ (in Planck units) and $\theta_{min}=93\pi$ using the  parameters in Table \ref{tab:7}. }
\label{tab:8}
\end{table}

\end{appendix}

\addcontentsline{toc}{section}{References}
\bibliographystyle{utphys}

\bibliography{refs}

\end{document}